\documentclass[aps,prb,reprint,superscriptaddress,amsmath,amssymb,floatfix]{revtex4-2}

\usepackage{graphicx}
\usepackage{dcolumn}
\usepackage{bm}           
\usepackage[colorlinks=true, linkcolor=blue, urlcolor=blue, citecolor=blue]{hyperref}
\usepackage[utf8]{inputenc}

\usepackage{xparse}

\begin{document}
\title{Symmetry-Protected Bipartite Entanglement in Hopf-Linked Quantum Rings}
\author{V. Yogesh}
\email{yo.physics@gmail.com}
\thanks{Corresponding author.}
\affiliation{Department of Physics, Government Arts College, Nandanam, Chennai - 600 035, India}

\author{Prosenjit Maity}
\email{prosenjit.maity@rkmrc.in}
\affiliation{Department of Physics, Ramakrishna Mission Residential College, Narendrapur, Kolkata - 700 103, India}

\begin{abstract}
We study two interacting electrons in Hopf-linked quantum rings, one electron on each ring. In the absence of the inter-ring interaction, the two lowest orbital states on each ring have the same energy. For the ideal geometry, the Fourier coefficients of the three-dimensional Coulomb kernel satisfy $V_{1,0}=-V_{0,1}$, making the positive-parity block of the four-state Hamiltonian diagonal in the Bell basis. Unitary symmetries of the full orbital Hamiltonian preserve the parity sectors under virtual transitions. For pure states odd under ring exchange followed by a half-turn on each ring ($\mathcal Y=-1$), the Schmidt weights occur in equal pairs, giving at least one bit of entanglement between the rings. An elliptic deformation changes the Coulomb selection rule to first order while preserving the link. Finite-basis diagonalization shows that higher orbitals can reduce the probability of obtaining the encoded Bell pair by projection onto the two lowest orbitals of each ring even when the entanglement between the rings remains close to one bit.
\end{abstract}
\maketitle

\section{Introduction}\label{sec:intro}
Spatial asymmetries in solid-state architectures modify kinetic degeneracies and the structure of entangled superpositions. Semiconductor quantum dots provide settings for spin qubits~\cite{Loss1998}. They exhibit orbital relaxation~\cite{Fujisawa2002}, and the coherence of charge qubits is sensitive to electrostatic noise~\cite{Petersson2010,PaqueletWuetz2023}. Quantum rings add sensitivity to gauge phases and geometry~\cite{Aharonov1959,Webb1985,Viefers2004}. For coupled electronic subsystems, the Coulomb interaction both generates correlations and changes the orbital eigenstates~\cite{Hichri2004}. These features motivate a model in which the relation between spatial geometry, discrete symmetry, and bipartite entanglement can be examined explicitly.

We introduce a minimal geometric model on a Hopf-linked configuration: two interacting electrons confined to spatially separated, non-coplanar one-dimensional (1D) quantum rings. The interaction is evaluated using its unexpanded three-dimensional (3D) distance dependence. A local Rashba spin--orbit coupling (SOC)~\cite{Bychkov1984,Meijer2002,Winkler2003,Bercioux2015}, together with an aligned exchange texture, supplies a possible construction of the required effective gauge field. Projecting onto the selected spin branch leaves the orbital motion on each ring as the dynamical degree of freedom. We take one electron on each ring and neglect inter-ring tunneling. The orbital states on the two rings then define the bipartition, while the two-electron state remains antisymmetric under particle exchange.

The Hopf linking number characterizes the embedding of the two centerlines. The electronic selection rules studied here follow from discrete transformations of the Hamiltonian. Changes in shape, local potentials, or effective flux can preserve the link and modify those transformations. The ideal-model block structure is protected by these symmetries, while its matrix elements vary with geometry.

The four-state calculation isolates two positive-parity Bell eigenstates and an internal geometric exchange gap. We develop this construction in two directions. First, an explicit symmetry representation establishes how virtual transitions enter a symmetry-respecting effective Hamiltonian. Second, orbital-basis convergence allows entanglement, projection weight, and spatial correlations to be evaluated beyond the two-qubit approximation. The resulting framework distinguishes an exact selection rule from an approximate low-energy encoding and from its sensitivity to a specified perturbation.

\section{Bipartite Geometry and Unperturbed Hamiltonian}\label{sec:geometry}
\begin{figure}[tbp]
\centering\includegraphics[width=.85\columnwidth]{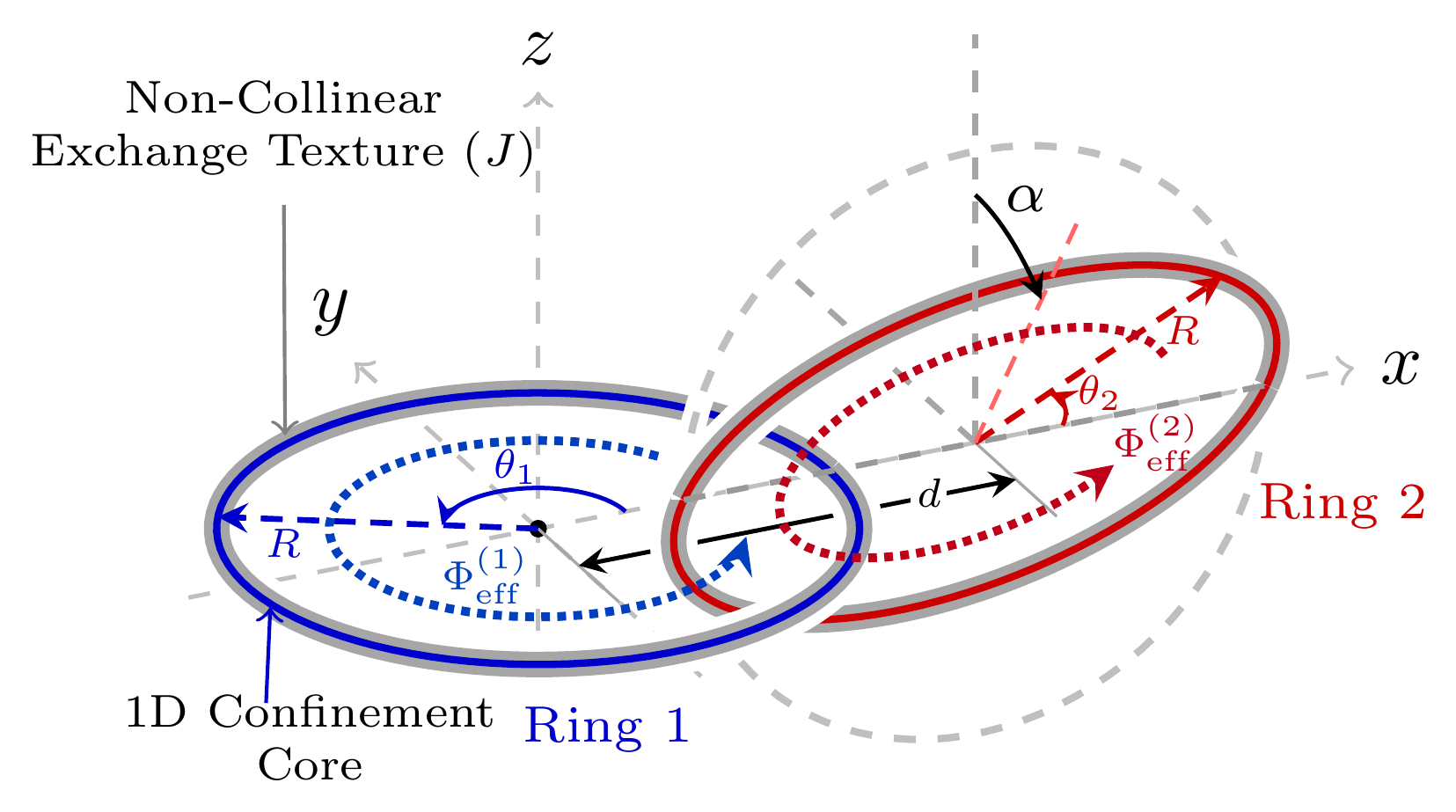}
\caption{Schematic of linked rings of radius $R$, with center
separation $d=\beta R$ and tilt $\alpha$ measured from perpendicular
ring planes. Dotted paths indicate effective gauge phases, and the
surrounding shells depict the conceptual local exchange arrangement.}
\label{fig:schematic}
\end{figure}
As illustrated in Fig.~\ref{fig:schematic}, we consider two interacting electrons restricted to 1D confinement cores. In a common Cartesian frame, their centerline coordinates are
\begin{align}
\mathbf r_1&=R(\cos\theta_1,\sin\theta_1,0),\label{eq:r1}\\
\mathbf r_2&=R(\beta+\cos\theta_2,-\sin\alpha\sin\theta_2,
\cos\alpha\sin\theta_2).\label{eq:r2}
\end{align}
Here $\theta_1,\theta_2$ are angular coordinates modulo $2\pi$ (in radians), $\beta=d/R$, and $\alpha=0$ denotes perpendicular planes. For $0<\beta<2$ and $|\alpha|<\pi/2$, the centerlines are disjoint and Hopf linked. The endpoints $\beta=0,2$ contain contacts. Finite transverse confinement introduces an additional clearance requirement: for a transverse radius $a_\perp$, a sufficient centerline criterion is
\begin{equation}
 R\min_{\theta_1,\theta_2}\mathcal D(\theta_1,\theta_2)>2a_\perp,
 \qquad \mathcal D=|\mathbf r_1-\mathbf r_2|/R.\label{eq:clearance}
\end{equation}
The contact condition follows from Eqs.~\eqref{eq:r1} and \eqref{eq:r2}: for $|\alpha|<\pi/2$, equality of the two positions requires $\sin\theta_2=0$, followed by $\sin\theta_1=0$. The remaining $x$ equation permits only $\beta=-2,0,2$. For $0<\beta<2$, ring 2 crosses the plane of the disk bounded by ring 1 at $x/R=\beta\pm1$. Exactly the point $\beta-1$ lies inside that disk, and the crossing is transverse. The oriented intersection number has magnitude one. Within this parameter domain the circles can be deformed continuously to the perpendicular configuration $(\beta,\alpha)=(1,0)$ without contact, so they retain the Hopf link type. The endpoints and the coplanar limit require separate treatment.

The numerical calculations sample $0.01\leq\beta\leq1.99$. For rings of finite transverse size, the sampled translations must also satisfy the clearance condition.

Projecting onto the targeted $\sigma=-1$ spin branch yields the effective orbital Hamiltonian
$H_0=\sum_{j=1}^2[\hbar^2/(2m^*R^2)](p_j-3/2)^2$, where $p_j=-i\partial_{\theta_j}$ is the dimensionless angular momentum on ring $j$, $m^*$ is the effective mass, and $\hbar$ is the reduced Planck constant. The orbitals $p_j=1,2$ have the same lowest energy, giving four unperturbed two-ring ground states. With $E_z=\hbar^2/(8m^*R^2)$ their energy is $E_0=2E_z$. Exciting one ring from either of these orbitals to $p_j=0$ or $3$ costs $\Delta E_{\mathrm{kin}}=8E_z$.
The Coulomb energy is $E_c=e^2/(4\pi\epsilon_0\epsilon_{\mathrm{eff}}R)$, where $e>0$ is the elementary charge, $\epsilon_0$ is the vacuum permittivity, and $\epsilon_{\mathrm{eff}}$ is the homogeneous effective relative permittivity. The interaction ratio is $\lambda=E_c/(8E_z)=R/a_B^*$, with effective Bohr radius $a_B^*=4\pi\epsilon_0\epsilon_{\mathrm{eff}}\hbar^2/(m^*e^2)$. The spin projection leaves a scalar orbital wavefunction. We write $H=H_0+V_{int}$ for the Hamiltonian on the full orbital space of this branch.

\subsection{Synthetic Gauge Field Initialization and Local Exchange}\label{app:gauge}
The four-state manifold requires degenerate one-electron orbitals $p=1$ and $p=2$ on each ring. To obtain separate effective Hamiltonians for the rings, we use a synthetic gauge field generated by a non-Abelian spin--orbit texture described by $SU(2)$, the group of two-dimensional unitary matrices with determinant one. Starting from the Rashba coupling $H_R=(\lambda_R/\hbar)(\bm\sigma\times\mathbf p)_z$, its Hermitian 1D form is $(\lambda_R/2R)\{\sigma_r(\theta_j),p_j\}$, where $\{A,B\}=AB+BA$ denotes the anticommutator and $\sigma_r=\sigma_x\cos\theta_j+\sigma_y\sin\theta_j$~\cite{Meijer2002}. The vector $\bm\sigma$ contains spin Pauli matrices, and $\mathbf p$ denotes physical momentum, distinct from the dimensionless $p_j$. These spin and momentum components are expressed in a local Cartesian frame whose $z$ axis is normal to the corresponding ring plane. The kinetic operator takes the form
\begin{equation}
 H_j^{kin}=\frac{\hbar^2}{2m^*R^2}
 \left(-i\frac{\partial}{\partial\theta_j}
 +\frac{m^*\lambda_RR}{\hbar^2}\sigma_r(\theta_j)\right)^2
 -\frac{m^*\lambda_R^2}{2\hbar^2}.\label{eq:rashba}
\end{equation}
To absorb the spin connection into a geometric phase, we apply a local non-Abelian $SU(2)$ gauge transformation to the spinor wavefunction:
\begin{equation}
 U(\theta_j)=e^{-i\theta_j\sigma_z/2}e^{i\gamma\sigma_y/2},
 \qquad \tan\gamma=\frac{2m^*\lambda_RR}{\hbar^2}.\label{eq:U}
\end{equation}
As derived in Appendix~\ref{app:spinsteps}, the transformed covariant momentum is $p_j-\tfrac12\sec\gamma\,\sigma_z$. Since $U(\theta_j+2\pi)=-U(\theta_j)$, the transformed spinor is antiperiodic when the laboratory spinor is periodic. Writing it as $e^{-i\theta_j/2}$ times a periodic orbital spinor supplies the half-integer shift in the effective flux.

The Rashba kinetic term alone is time-reversal invariant. A uniform axial Zeeman term becomes proportional to $\sin\gamma\,\sigma_x+\cos\gamma\,\sigma_z$ in the rotated frame and generally mixes its spin branches. We choose an aligned exchange texture and retain the Rashba and exchange terms in the ideal spin Hamiltonian.

To isolate the desired branch, take $J>0$ and an exchange field aligned with the rotated spin axis:
\begin{equation}
 H_{ex}^{local}=-JU\sigma_zU^\dagger
 =-J\cos\gamma\,\sigma_z+J\sin\gamma\,\sigma_r.\label{eq:exchange}
\end{equation}
The target effective flux is $\Phi_{\mathrm{eff}}^{(j)}/\Phi_0=3/2$, where $\Phi_0=h/e$ is the flux quantum and $h=2\pi\hbar$. This value is obtained by setting $\lambda_R=\sqrt3\hbar^2/(2m^*R)$, which fixes $\gamma=\pi/3$. Substituting this requirement gives the conically canted local exchange field and its rotating-frame form:
\begin{align}
H_{ex}^{local}&=-\frac J2\sigma_z+\frac{\sqrt3J}{2}\sigma_r(\theta_j),\label{eq:conicalfield}\\
H_{ex}^{eff}&=U^\dagger H_{ex}^{local}U=-J\sigma_z.\label{eq:rotatedexchange}
\end{align}
Denoting the rotated $\sigma_z$ eigenvalue by $-\sigma$, with $\sigma=\pm1$, gives the branch Hamiltonian
\begin{equation}
H_j^{eff,\sigma}=4E_z\left(p_j-\frac{\Phi_{\mathrm{eff}}^{(j,\sigma)}}{\Phi_0}\right)^2+\sigma J-\frac{m^*\lambda_R^2}{2\hbar^2}.\label{eq:spinbranch}
\end{equation}
The effective flux is
\begin{equation}
\frac{\Phi_{\mathrm{eff}}^{(j,\sigma)}}{\Phi_0}=\frac12\left[1-\sigma\sqrt{1+\left(\frac{2m^*\lambda_RR}{\hbar^2}\right)^2}\right].\label{eq:flux}
\end{equation}
The resulting flux is $\Phi_{\mathrm{eff}}^{(j,-)}/\Phi_0=3/2$. The kinetic minima of the two branches are equal; exchange lowers the $\sigma=-1$ minimum by $2J$ relative to the conjugate branch, whose flux is $\Phi_{\mathrm{eff}}^{(j,+)}/\Phi_0=-1/2$. Common offsets are absorbed into the orbital energy reference in $H_0$. A sufficiently large splitting compared with spin-mixing perturbations is required for spin projection. The aligned exchange term has angular average $-J\cos\gamma\,\sigma_z$ and therefore a finite axial component. The orbital model assumes these gauge fields and the aligned texture; a material implementation must also control stray fields (Sec.~\ref{sec:implementation}).

\subsection{Ring Modes and Fermionic Antisymmetry}\label{sec:bipartition}
Let $c_{jp}^\dagger$ create an electron in orbital $p$ on ring $j$. With one electron on each ring, we fix the order of the creation operators and identify the two-ring orbital basis as
\begin{equation}
 |p,q\rangle=c_{1p}^\dagger c_{2q}^\dagger|0\rangle
 \longleftrightarrow |p\rangle_1\otimes|q\rangle_2.\label{eq:modes}
\end{equation}
Here $p$ and $q$ label angular-momentum orbitals on rings 1 and 2, respectively, and $|0\rangle$ denotes the fermionic vacuum. The creation operators anticommute, preserving fermionic antisymmetry. The tensor factors label the rings~\cite{Zanardi2002}; the two orbitals $\{|1\rangle_j,|2\rangle_j\}$ on each ring encode its qubit. If tunneling transfers an electron between the rings, either ring can be empty or doubly occupied. The rings still define the bipartition, but their joint state space now includes $(2,0)$ and $(0,2)$ as well as $(1,1)$, where the pair gives the electron numbers on rings 1 and 2. Section~\ref{app:tunneling} gives the condition for neglecting these charge-transfer states.

\section{Non-Perturbative Coulomb Matrix Elements}
\label{sec:coulomb}

In the interlocked regime ($0<\beta<2$), the standard center-based Legendre multipole expansion encounters spatial configurations outside its convergence radius. Its uniform application over the full angular integration domain requires $d>2R$, or $\beta>2$. We therefore use matrix elements of the unexpanded Coulomb interaction to construct the Hamiltonian, $H=H_0+V_{int}$. This geometric restriction concerns the spatial multipole expansion; a perturbative treatment in the interaction strength remains governed by the coupling relative to the relevant spectral separations.

The interaction is given by $V_{int}=E_c[\mathcal D^2(\theta_1,\theta_2;\alpha,\beta)]^{-1/2}$, with characteristic energy $E_c=e^2/(4\pi\epsilon_0\epsilon_{\mathrm{eff}}R)$. The one-dimensional centerline model neglects transverse wavefunction spreading. From the ring coordinates in Eqs.~\eqref{eq:r1} and \eqref{eq:r2}, the dimensionless displacement $\mathbf D=(\mathbf r_1-\mathbf r_2)/R$ has components
\begin{align}
D_x&=\cos\theta_1-\beta-\cos\theta_2,\nonumber\\
D_y&=\sin\theta_1+\sin\alpha\sin\theta_2,\nonumber\\
D_z&=-\cos\alpha\sin\theta_2.\label{eq:components}
\end{align}
Squaring and adding these components, using $\sin^2\theta_j+\cos^2\theta_j=1$ and $\sin^2\alpha+\cos^2\alpha=1$, gives $\mathcal D^2=\mathcal D_s^2+2\sin\alpha\sin\theta_1\sin\theta_2$, where $\mathcal D_s^2=2+\beta^2-2\cos\theta_1\cos\theta_2-2\beta(\cos\theta_1-\cos\theta_2)$ is the squared distance at $\alpha=0$. Here $\alpha=0$ corresponds to orthogonal ring planes.

For the separate softening tests, we use $V_{int}^{(\rho)}=E_c/\sqrt{\mathcal D^2+\rho_{\mathrm{soft}}^2}$. The dimensionless softening parameter $\rho_{\mathrm{soft}}$ is distinct from the transverse geometric radius entering the clearance condition in Eq.~\eqref{eq:clearance}. A finite-width interaction would require averaging the Coulomb kernel over the transverse wavefunctions. The calculations below use these centerline and softened kernels. In a finite-width ring, entanglement also depends on the transverse confinement and the resulting averaged interaction.

Projecting the interaction potential onto the unperturbed angular momentum basis yields the spatial matrix elements $V_{m,n}(\alpha,\beta)$, hereafter denoted $V_{m,n}$ for brevity. The single-particle angular functions are $\psi_p(\theta)=(2\pi)^{-1/2}e^{ip\theta}$. We define the Coulomb Fourier coefficients by
\begin{equation}
V_{m,n}=\frac{E_c}{(2\pi)^2}\int_{-\pi}^{\pi}\!d\theta_1\int_{-\pi}^{\pi}\!d\theta_2\,
\frac{e^{i(m\theta_1+n\theta_2)}}{\mathcal D(\theta_1,\theta_2;\alpha,\beta)}.\label{eq:fourier}
\end{equation}
With this positive-phase convention, $\langle p_1,p_2|V_{int}|p'_1,p'_2\rangle=V_{p'_1-p_1,p'_2-p_2}$. The integers $m,n$ are the corresponding momentum transfers. Simultaneous reversal of both transfers leaves the coefficients unchanged for the present geometry, as established in Sec.~\ref{app:symmetry}.

\begin{figure}[tbp]\centering
\includegraphics[width=0.8\columnwidth]{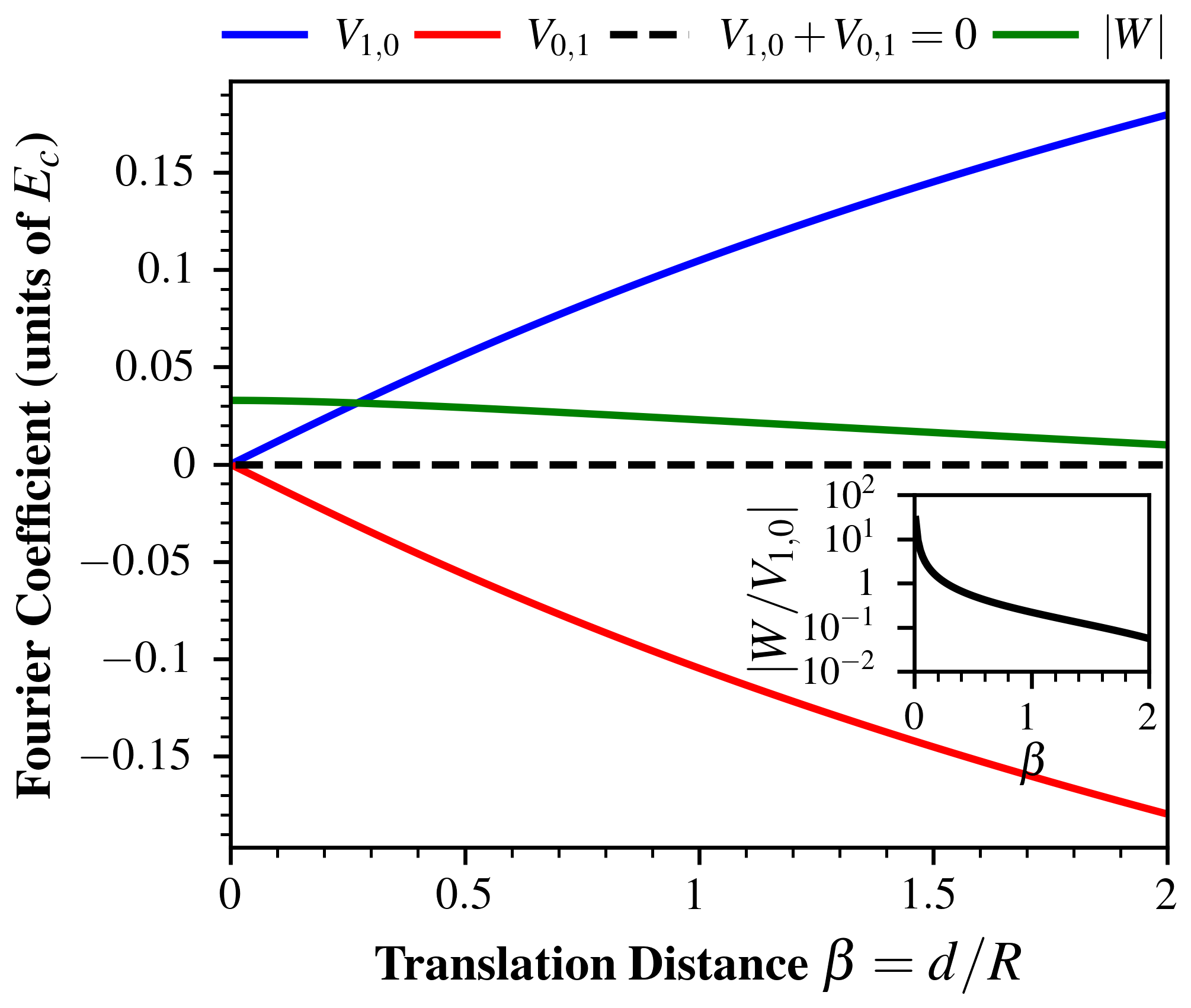}
\caption{Coulomb Fourier coefficients $V_{1,0}/E_c$, $V_{0,1}/E_c$, their sum, and $|W|/E_c$ as functions of translation distance $\beta=d/R$, with $W=\tfrac12(V_{1,1}-V_{1,-1})$. The tilt is $\alpha=0.3$ rad and the interaction is unsoftened, $\rho_{\mathrm{soft}}=0$. The calculation uses 101 uniformly spaced translations over $0.01\leq\beta\leq1.99$. Periodic angular quadrature is refined through $N=300,600,1200,2400,4800$ until the coefficient criterion in Appendix~\ref{app:numerics} is satisfied. The inset shows $|W/V_{1,0}|$ on a logarithmic scale. The vanishing sum tests the numerical implementation of the analytical identity $V_{1,0}=-V_{0,1}$.}
\label{fig:symmetry}\end{figure}

The unexpanded spatial distance is invariant under the discrete $\mathbb Z_2$ parity-exchange bijection $\Pi_{12}:(\theta_1,\theta_2)\mapsto(\pi-\theta_2,\pi-\theta_1)$. Together with simultaneous angular inversion, this symmetry enforces the identity $V_{1,0}=-V_{0,1}$. Figure~\ref{fig:symmetry} illustrates this cancellation and the geometric difference $W=\tfrac12(V_{1,1}-V_{1,-1})$.

The inset of Fig.~\ref{fig:symmetry} shows the ratio $|W/V_{1,0}|$ of the geometric exchange scale to the magnitude of the single-transfer coefficient. Appendix~\ref{app:numerics} gives the angular quadrature and its convergence criterion. The spectral roles of these coefficients follow from the projected Hamiltonian in Sec.~\ref{sec:parity}.

\subsection{Multipole Expansion in the Near Field}\label{sec:multipole_main}
A standard multipole treatment of the inter-ring Coulomb interaction expands the inverse distance about the geometric centers of the two rings. Let the vector separating the centers be $\mathbf d$, where $|\mathbf d|=d=\beta R$. Let $\mathbf r_1$ denote the position on ring 1 relative to its center, and $\mathbf r'_2$ the position on ring 2 relative to its center. The Coulomb distance is then $|\mathbf d+\mathbf r'_2-\mathbf r_1|$. The center-based Legendre expansion is~\cite{Jackson1998}
\begin{equation}
\begin{aligned}
&\frac{1}{|\mathbf d+(\mathbf r'_2-\mathbf r_1)|}\\
&\qquad=\frac1d\sum_{\ell=0}^{\infty}P_\ell(\cos\Theta)\left(\frac{|\mathbf r_1-\mathbf r'_2|}{d}\right)^\ell,
\end{aligned}\label{eq:multipole}
\end{equation}
where $P_\ell$ is the Legendre polynomial of degree $\ell$ and $\Theta$ is the angle between $\mathbf d$ and $\mathbf r_1-\mathbf r'_2$. The index $\ell$ distinguishes the multipole order from the Fourier transfer index $n$.

For expansion ratios strictly below unity, $|P_\ell(\cos\Theta)|\leq1$ bounds the series by a convergent geometric series. Absolute and uniform convergence over the angular domain therefore follows from
\begin{equation}
\max_{\theta_1,\theta_2}|\mathbf r_1-\mathbf r'_2|<d.\label{eq:equation2}
\end{equation}
The triangle inequality gives $|\mathbf r_1-\mathbf r'_2|\leq|\mathbf r_1|+|\mathbf r'_2|=2R$. For the present rings, equality is attained at opposite points on their common local $x$ axis, for example $\theta_1=0$ and $\theta_2=\pi$. Thus the maximum is exactly $2R$, and the strict uniform-convergence condition is $\beta>2$.

For $0<\beta<2$, the expansion ratio exceeds unity at these configurations and throughout a surrounding open set. The center-based series is therefore unsuitable for the full angular integral, which we evaluate with the unexpanded kernel. Appendix~\ref{app:multipole} gives the generating-function derivation and the convergence bound. The resulting Fourier coefficients can be used in either a weak-coupling calculation or a diagonalization in a specified orbital basis.

\subsection{Exact Symmetry Proof: \texorpdfstring{$V_{1,0}=-V_{0,1}$}{V(1,0)=-V(0,1)}}\label{app:symmetry}
The identity follows directly from the geometry and holds for arbitrary integer momentum transfers in the full orbital problem.

First, the real Coulomb kernel gives $V_{m,n}^*=V_{-m,-n}$. The squared distance is also invariant under simultaneous angular inversion, $(\theta_1,\theta_2)\mapsto(-\theta_1,-\theta_2)$: the cosine terms are unchanged and both sines change sign. Substitution in Eq.~\eqref{eq:fourier} therefore gives $V_{-m,-n}=V_{m,n}$. Combining these relations establishes
\begin{equation}
V_{m,n}^*=V_{-m,-n}=V_{m,n}.\label{eq:equation3}
\end{equation}
Thus all the Fourier coefficients are real for this kernel.

We next apply the bijective parity-exchange transformation
\begin{equation}
\theta_1=\pi-\phi_2,\qquad\theta_2=\pi-\phi_1,\label{eq:equation4}
\end{equation}
with angular coordinates understood modulo $2\pi$. The elementary trigonometric identities give
\begin{align}
\cos\theta_1\cos\theta_2&=\cos\phi_1\cos\phi_2,\nonumber\\
\cos\theta_1-\cos\theta_2&=\cos\phi_1-\cos\phi_2,\nonumber\\
\sin\theta_1\sin\theta_2&=\sin\phi_1\sin\phi_2.\label{eq:equation5}
\end{align}
Each term in $\mathcal D_s^2$ and the tilt-dependent term is therefore invariant separately. Consequently, $\mathcal D^2(\pi-\phi_2,\pi-\phi_1;\alpha,\beta)=\mathcal D^2(\phi_1,\phi_2;\alpha,\beta)$.

In the disjoint centerline domain $0<\beta<2$ and $|\alpha|<\pi/2$, the distance is positive. Taking its positive square root yields
\begin{equation}
\mathcal D(\pi-\phi_2,\pi-\phi_1;\alpha,\beta)=\mathcal D(\phi_1,\phi_2;\alpha,\beta).\label{eq:distanceinvariance}
\end{equation}
The absolute Jacobian of the transformation is one. Periodicity permits the transformed integration domain to be shifted back to $[-\pi,\pi]^2$.

For integer $m,n$, the numerator transforms as $e^{i(m\theta_1+n\theta_2)}=(-1)^{m+n}e^{-i(n\phi_1+m\phi_2)}$. Substitution in Eq.~\eqref{eq:fourier}, followed by Eq.~\eqref{eq:equation3}, gives the general selection rule
\begin{equation}
\begin{aligned}V_{m,n}&=(-1)^{m+n}V_{-n,-m}\\&=(-1)^{m+n}V_{n,m}.
\end{aligned}\label{eq:selection}
\end{equation}
For the single-transfer coefficient, the cancellation can be written explicitly:
\begin{equation}
\begin{aligned}
V_{1,0}&=-\frac{E_c}{(2\pi)^2}\int_{-\pi}^{\pi}\!d\phi_1\int_{-\pi}^{\pi}\!d\phi_2\,
\frac{e^{-i\phi_2}}{\mathcal D(\phi_1,\phi_2)}\\
&=-V_{0,-1}=-V_{0,1}.
\end{aligned}\label{eq:singletransferproof}
\end{equation}
The identity is exact throughout the stated disjoint domain and uses the unexpanded interaction.

A constant softening term $\rho_{\mathrm{soft}}^2$ preserves both coordinate transformations, so the same selection rule holds for the softened kernel. More generally, the argument applies to any integrable potential that depends on the angles only through $\mathcal D^2$. Homogeneous dielectric screening changes the overall scale $E_c$ and preserves the identity.

Screening by a metallic gate depends on each charge's position relative to it. The image-charge separation generally differs from $R\sqrt{\mathcal D^2+(2h/R)^2}$ for a noncoplanar geometry, where $h$ denotes a proposed gate height. The selection rule survives such screening when the complete electrostatic kernel is invariant under both simultaneous angular inversion and parity exchange. Gate-induced one-body potentials must also respect the corresponding symmetries for the full Hamiltonian to retain them. Both conditions depend on the gate geometry.

\section{Parity Manifolds and the Geometric Exchange Gap}\label{sec:parity}
\begin{figure}[tbp]\centering
\includegraphics[width=\columnwidth]{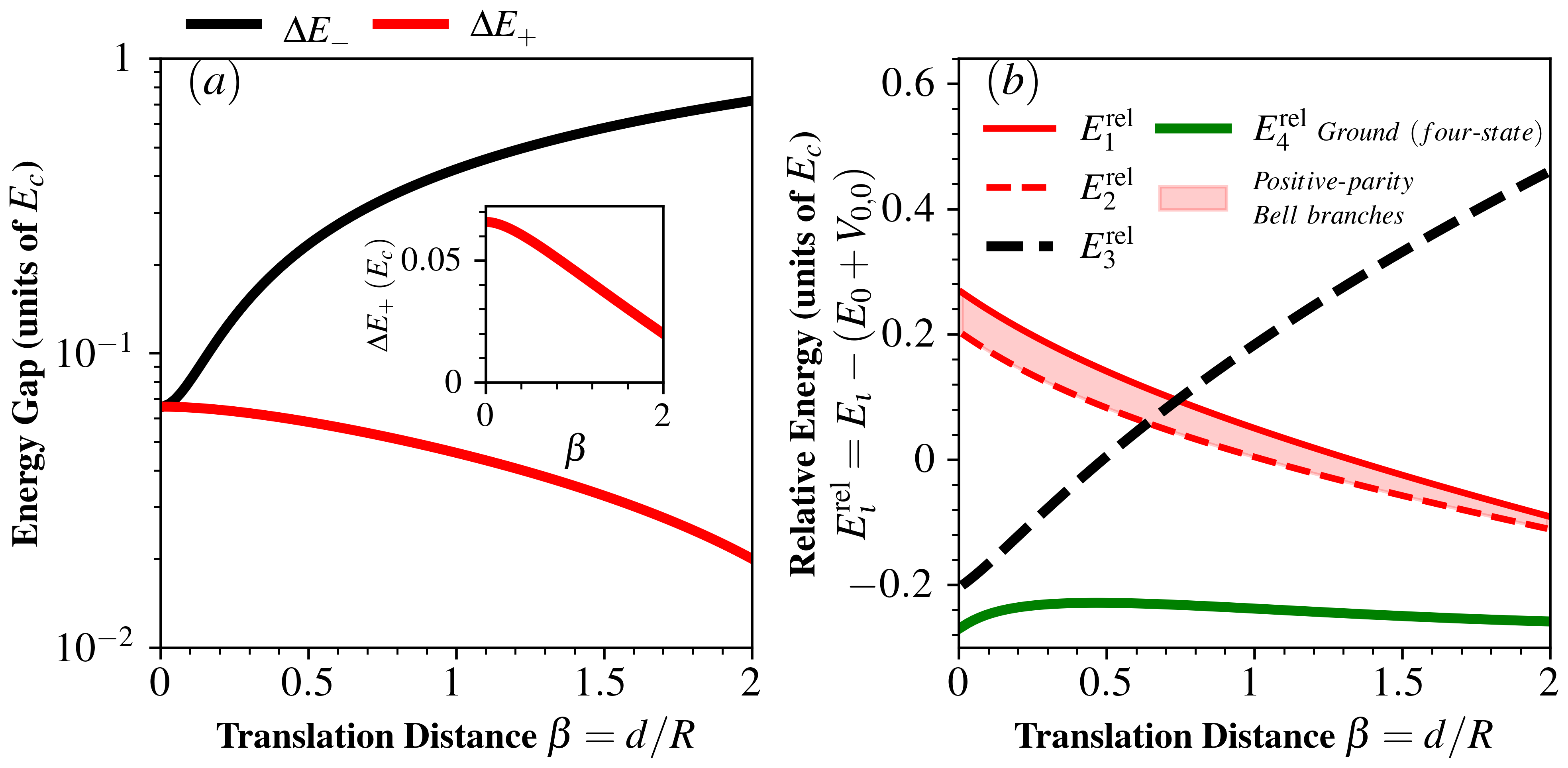}
\caption{Four-state spectrum at a fixed geometric tilt $\alpha=0.3$ rad, with an unsoftened interaction ($\rho_{\mathrm{soft}}=0$). The calculation uses 101 uniformly spaced translations $0.01\leq\beta\leq1.99$ and the angular quadrature refinement specified in Fig.~\ref{fig:symmetry} and Appendix~\ref{app:numerics}. (a) Internal parity-sector splittings $\Delta E_\pm/E_c$ on a logarithmic scale. The inset shows $\Delta E_+/E_c$ on a linear scale. (b) The four eigenvalues relative to $E_0+V_{0,0}$, in units of $E_c$. The red branches are the positive-parity Bell levels; the remaining branches are the eigenvalues of $H^-$. The shaded interval indicates the internal separation of the positive-parity levels. The lowest displayed level is the ground state within the four-state truncation.}
\label{fig:energy_analysis}\end{figure}

Having established the exact spatial symmetry of the Coulomb interaction, $V_{1,0}=-V_{0,1}$, we now evaluate its effect on the unperturbed bipartite ground-state manifold. We transform the standard product basis into the maximally entangled Bell basis, sorted by the eigenvalues of the orbital pseudospin parity operator $\Sigma_x=\sigma_x^{(1)}\otimes\sigma_x^{(2)}$. Here the Pauli matrices act on the orbital states $p_j\in\{1,2\}$, separately from the physical spin matrices used in the gauge-field construction.

This transformation separates the projected Hamiltonian into positive- and negative-parity blocks. The spatial identity cancels the off-diagonal coupling between the two positive-parity Bell states, while the corresponding negative-parity coupling remains finite. The geometric exchange energy $W(\alpha,\beta)\equiv W=\tfrac12(V_{1,1}-V_{1,-1})$ determines the internal positive-parity splitting $\Delta E_+=2|W|$, shown in Fig.~\ref{fig:energy_analysis}. A finite splitting is compatible with exact parity symmetry: it separates two states with the same parity.

\subsection{Bell Basis Block-Diagonalization}\label{app:bell}
In the ordered product basis $\mathcal P=\{|1,1\rangle,|1,2\rangle,|2,1\rangle,|2,2\rangle\}$, the projected Hamiltonian is
\begin{equation}
H_{\mathcal P}=E_0\mathbb I+
\begin{pmatrix}
V_{0,0}&V_{0,1}&V_{1,0}&V_{1,1}\\
V_{0,-1}&V_{0,0}&V_{1,-1}&V_{1,0}\\
V_{-1,0}&V_{-1,1}&V_{0,0}&V_{0,1}\\
V_{-1,-1}&V_{-1,0}&V_{0,-1}&V_{0,0}
\end{pmatrix},\label{eq:HP}
\end{equation}
where $\mathbb I$ is the identity on $\mathcal P$. The inversion identity in Eq.~\eqref{eq:equation3} makes this matrix real and symmetric.

The orbital pseudospin parity operator has the representation
\begin{equation}
\Sigma_x=\begin{pmatrix}0&0&0&1\\0&0&1&0\\0&1&0&0\\1&0&0&0\end{pmatrix}.\label{eq:sigmax_main}
\end{equation}
Simultaneous reversal of both Fourier transfers gives $[H_{\mathcal P},\Sigma_x]=0$. Here $[A,B]=AB-BA$ denotes the commutator. The parity eigenstates are the Bell pairs
\begin{align}
|\Psi_1^\pm\rangle&=\frac{|1,1\rangle\pm|2,2\rangle}{\sqrt2},\label{eq:equation6}\\
|\Psi_2^\pm\rangle&=\frac{|1,2\rangle\pm|2,1\rangle}{\sqrt2}.\label{eq:Bell}
\end{align}
They satisfy $\Sigma_x|\Psi_i^\pm\rangle=\pm|\Psi_i^\pm\rangle$. These parity labels refer to orbital-state reversal within the encoding.

In the ordered Bell basis $\{|\Psi_1^+\rangle,|\Psi_2^+\rangle,|\Psi_1^-\rangle,|\Psi_2^-\rangle\}$, the Hamiltonian takes the form $E_0\mathbb I+H^+\oplus H^-$, where $H^\pm(\alpha,\beta)\equiv H^\pm$. Evaluating the matrix elements gives
\begin{equation}
H^\pm=\begin{pmatrix}
V_{0,0}\pm V_{1,1}&\pm V_{1,0}+V_{0,1}\\
\pm V_{1,0}+V_{0,1}&V_{0,0}\pm V_{1,-1}
\end{pmatrix}.\label{eq:blocks}
\end{equation}
For example, the positive-parity off-diagonal element is
\[
\begin{aligned}
\langle\Psi_1^+|V_{int}|\Psi_2^+\rangle
&=\tfrac12(V_{0,1}+V_{1,0}+V_{-1,0}+V_{0,-1})\\
&=V_{0,1}+V_{1,0}.
\end{aligned}
\]
The corresponding negative-parity element is $V_{0,1}-V_{1,0}$. Opposite-parity matrix elements vanish because the Hamiltonian commutes with $\Sigma_x$. Appendix~\ref{apx:bell_spectra} gives the explicit basis-change matrix and the characteristic polynomial of the negative-parity block.

Substituting $V_{1,0}+V_{0,1}=0$ makes $H^+$ diagonal. Its eigenstates are therefore the individual positive-parity Bell states, with energies $E_0+V_{0,0}+V_{1,1}$ and $E_0+V_{0,0}+V_{1,-1}$. Their separation is $\Delta E_+=|V_{1,1}-V_{1,-1}|=2|W|$.

For $H^-$, the same identity gives an off-diagonal element $-2V_{1,0}$. Subtracting the mean of its diagonal entries leaves diagonal elements $-W,W$. The two eigenvalues of this traceless block differ by $\Delta E_-=2\sqrt{W^2+4V_{1,0}^2}$.

Relative to the common energy $E_0+V_{0,0}$, the positive-parity levels are $V_{1,1}$ and $V_{1,-1}$, while the negative-parity levels are
\begin{equation}
-\frac{V_{1,1}+V_{1,-1}}2\pm\sqrt{W^2+4V_{1,0}^2}.\label{eq:equation7}
\end{equation}
These expressions give the four branches in Fig.~\ref{fig:energy_analysis}(b).

The quantities $\Delta E_\pm$ describe separations within their respective parity sectors. Isolation of an individual Bell level also depends on its separation from the other sector and from higher orbitals. The coupling $-2V_{1,0}$ produces level repulsion within the negative-parity block; spatial localization requires an independent examination of the eigenstates.

The gap formulas imply $\Delta E_-\geq\Delta E_+$ and $0\leq\Delta E_--\Delta E_+\leq4|V_{1,0}|$. Thus the two internal splittings approach one another as $V_{1,0}$ tends to zero. This relation accounts for the approach of the two internal splittings at small positive $\beta$ in Fig.~\ref{fig:energy_analysis}(a). The contacting geometry at $\beta=0$ lies outside the disjoint centerline domain considered here.

The eigenstates of the parity blocks also determine their orbital concurrence. A normalized state in either block can be written as $|\chi^\pm\rangle=c_1|\Psi_1^\pm\rangle+c_2|\Psi_2^\pm\rangle$, with $|c_1|^2+|c_2|^2=1$. Transforming to the product basis and applying the pure-state concurrence formula~\cite{Wootters1998} gives $\mathcal C_\pm=|c_1^2-c_2^2|$. This expression remains valid for complex amplitudes.

For $W\neq0$, the positive-parity eigenvectors are the separate Bell states, and their concurrence is
\begin{equation}
\begin{aligned}
\mathcal C_+&=\frac{|(V_{0,0}+V_{1,1})-(V_{0,0}+V_{1,-1})|}{\Delta E_+}\\
&=\frac{2|W|}{2|W|}=1.
\end{aligned}\label{eq:Cplus_parity}
\end{equation}
At $W=0$, the positive block is degenerate. Bell states remain possible eigenstates, but arbitrary superpositions in the same eigenspace can have different concurrences. A state-selection prescription is then required.

For a nondegenerate eigenstate of the real symmetric negative-parity block, its coefficients can be chosen real up to an overall phase. Diagonalization gives $|c_1^2-c_2^2|=|W|/\sqrt{W^2+4V_{1,0}^2}$, and hence
\begin{equation}
\begin{aligned}
\mathcal C_-&=\frac{|(V_{0,0}-V_{1,1})-(V_{0,0}-V_{1,-1})|}{\Delta E_-}\\
&=\frac{|W|}{\sqrt{W^2+4V_{1,0}^2}}.
\end{aligned}\label{eq:Cminus_parity}
\end{equation}
This result applies when $W^2+4V_{1,0}^2>0$. If both coefficients vanish, the negative block is also degenerate and its concurrence depends on the chosen eigenvector (Appendix~\ref{apx:bell_concurrence}).

The negative-parity concurrence is therefore controlled by the relative magnitudes of $W$ and $V_{1,0}$. It becomes small when $|V_{1,0}|$ dominates $|W|$, whereas it approaches unity when $|V_{1,0}|/|W|$ tends to zero. Section~\ref{sec:concurrence} examines this dependence on geometry.

The four-state projection establishes the Bell eigenvectors and internal splittings within the lowest unperturbed orbital shell. It identifies the two-qubit encoding and provides closed-form expressions for its spectrum and concurrence. The full Coulomb interaction also couples this shell to higher angular momentum states. These couplings can shift energies and reduce the weight of an eigenstate in $\mathcal P$. Their effect on parity is determined by the symmetries of the enlarged Hamiltonian. Section~\ref{sec:dynamical_mixing} examines whether higher-orbital corrections can mix opposite parities.

\section{Higher-Orbital Mixing and Symmetry}\label{sec:dynamical_mixing}
The spatial parity-exchange bijection $\Pi_{12}$ preserves the Coulomb interaction but reverses and exchanges the canonical angular momenta. Consequently, $\Pi_{12}$ alone fails to commute with the kinetic Hamiltonian in the presence of the synthetic gauge shift. To determine whether the parity decomposition survives coupling to excited momentum branches, we examine the symmetries of the full spin-projected orbital Hamiltonian.

The Fourier selection rule $V_{m,n}=(-1)^{m+n}V_{n,m}$ permits nonzero cross-shell coefficients, including $V_{1,2}=-V_{2,1}$. These coefficients can admix higher orbitals. Which effective couplings survive depends on how the retained and virtual states transform under the symmetries of the Hamiltonian.

The higher-orbital correction can be organized using the Schrieffer--Wolff (SW) transformation~\cite{Schrieffer1966,Bravyi2011}. At second order, a virtual state $|d\rangle$ contributes a product $\langle a|V_{int}|d\rangle\langle d|V_{int}|b\rangle$ to an effective matrix element between states $|a\rangle$ and $|b\rangle$. For an off-diagonal matrix element, this product retains the relative signs of the two amplitudes even when all Fourier coefficients are real. The symmetry-related virtual contributions must therefore be combined before assessing parity mixing.

\subsection{Unitary Parity and Generalized Antiunitary Symmetry}\label{app:magnetic_group}
We work in the spin-projected orbital Hilbert space of periodic functions of $\theta_1$ and $\theta_2$. The canonical angular momentum operators are $p_j=-i\partial_{\theta_j}$, and the kinetic Hamiltonian is $H_0=4E_z[(p_1-\tfrac32)^2+(p_2-\tfrac32)^2]$.

For a periodic orbital wavefunction $f(\theta_1,\theta_2)$, simultaneous angular inversion $I$ acts as $(If)(\theta_1,\theta_2)=f(-\theta_1,-\theta_2)$. The parity-exchange operation acts as $(\Pi_{12}f)(\theta_1,\theta_2)=f(\pi-\theta_2,\pi-\theta_1)$, with angles understood modulo $2\pi$. On the periodic domain of Appendix~\ref{apx:domains}, their momentum transformations are $Ip_jI^{-1}=-p_j$ and $\Pi_{12}p_1\Pi_{12}^{-1}=-p_2$, $\Pi_{12}p_2\Pi_{12}^{-1}=-p_1$. These transformations reverse the effective gauge shifts in the kinetic energy.

On either ring the Fourier mode $e^{ip\theta_j}$ has integer angular-momentum index $p$ and kinetic energy $4E_z(p-3/2)^2$. A second index $p'$ has the same energy precisely when
\[
 (p'-\tfrac32)^2-(p-\tfrac32)^2
   =(p'-p)(p'+p-3)=0.
\]
Besides $p'=p$, the solution is $p'=3-p$. It reverses the shifted momentum, $p'-3/2=-(p-3/2)$; in particular, it interchanges the lowest orbitals $1$ and $2$, whose shifted momenta are $-1/2$ and $+1/2$.

This reflection can be implemented on each Fourier factor in two steps:
\[
 e^{ip\theta_j}\xrightarrow{\ I\ }e^{-ip\theta_j}
 \xrightarrow{\ \times\,e^{3i\theta_j}\ }e^{i(3-p)\theta_j}.
\]
Applying the two steps to both rings gives the full orbital parity operator,
\begin{equation}
\begin{aligned}
\mathcal X&=e^{3i(\theta_1+\theta_2)}I,\\
\mathcal X|p,q\rangle&=|3-p,3-q\rangle.
\end{aligned}\label{eq:X}
\end{equation}
Under a full turn of either ring the phase multiplier gains $e^{6\pi i}=1$, so $\mathcal X$ preserves periodic boundary conditions. Inversion takes the multiplier to its inverse, giving $\mathcal X^2=1$. The basis action gives $\mathcal Xp_j\mathcal X^{-1}=3-p_j$; hence $(p_j-3/2)^2$ and therefore $H_0$ are invariant.

The interaction is a real multiplicative function invariant under $I$. It commutes with the phase factor in $\mathcal X$, and hence is also invariant under $\mathcal X$. This establishes a unitary parity symmetry of the full orbital Hamiltonian.

A second useful unitary operation is
\begin{equation}
\begin{aligned}
\mathcal Y&=I\Pi_{12},\\
\mathcal Y|p,q\rangle&=(-1)^{p+q}|q,p\rangle.
\end{aligned}\label{eq:Y}
\end{equation}
Its coordinate action is $(\mathcal Yf)(\theta_1,\theta_2)=f(\theta_2+\pi,\theta_1+\pi)$. Thus $\mathcal Y$ exchanges the two angular momenta. The equal-ring kinetic energy is invariant under this exchange, and the interaction is invariant under both $I$ and $\Pi_{12}$. Furthermore, $\mathcal Y^2=1$ and $[\mathcal X,\mathcal Y]=0$. Consequently,
\begin{equation}
[H,\mathcal X]=[H,\mathcal Y]=0,\qquad\mathcal X|_{\mathcal P}=\Sigma_x.\label{eq:commutators}
\end{equation}
The restriction to $\mathcal P$ is well defined because $\mathcal X$ maps this four-state space into itself. The two positive-parity Bell states carry distinct $\mathcal Y$ eigenvalues: $\mathcal Y|\Psi_1^+\rangle=|\Psi_1^+\rangle$ and $\mathcal Y|\Psi_2^+\rangle=-|\Psi_2^+\rangle$. These labels distinguish their continuations into the enlarged orbital space; Appendix~\ref{apx:joint_sectors} gives the joint-sector projectors. Each state can acquire higher-orbital components within its allowed symmetry sector.

The orbital Hamiltonian also admits a generalized antiunitary symmetry. Let $K_c$ denote coordinate-space complex conjugation, $(K_cf)(\theta_1,\theta_2)=f(\theta_1,\theta_2)^*$, and define $\mathcal K_{\mathrm{global}}^{\mathrm{orb}}=K_c\Pi_{12}$. Its action on an orbital expansion is
\begin{equation}
\begin{aligned}
&\mathcal K_{\mathrm{global}}^{\mathrm{orb}}\sum_{p,q}c_{pq}|p,q\rangle\\
&\qquad=\sum_{p,q}c_{pq}^*(-1)^{p+q}|q,p\rangle.
\end{aligned}\label{eq:antiunitary}
\end{equation}
To obtain this result, $\Pi_{12}$ first maps $|p,q\rangle$ to $(-1)^{p+q}|-q,-p\rangle$, and $K_c$ reverses both orbital phases while conjugating the expansion coefficients.

Complex conjugation transforms $p_j$ to $-p_j$. Combined with the momentum reversal under $\Pi_{12}$, the antiunitary operation therefore exchanges $p_1$ and $p_2$ without changing their signs. It preserves the equal-ring kinetic Hamiltonian, including the identical gauge shifts. The Coulomb interaction is invariant because it is real and unchanged by $\Pi_{12}$. Thus
\begin{equation}
\mathcal K_{\mathrm{global}}^{\mathrm{orb}}H
\bigl(\mathcal K_{\mathrm{global}}^{\mathrm{orb}}\bigr)^{-1}=H.\label{eq:equation9}
\end{equation}
The coordinate transformation $\Pi_{12}$ is real and commutes with $K_c$. Since $K_c^2=\Pi_{12}^2=1$, the combined operation satisfies $(\mathcal K_{\mathrm{global}}^{\mathrm{orb}})^2=1$. This antiunitary operation squares to the identity and does not enforce Kramers degeneracy. The unitary operation $\mathcal X$ provides the parity label used in the orbital decomposition.

These symmetries act within the spin branch selected in Sec.~\ref{app:gauge}. Extending them to the Hamiltonian before projection requires the local exchange texture and the other spin-dependent terms to transform consistently. Appendix~\ref{app:operatorsteps} gives the coordinate actions, operator domains, and restrictions to the encoding. Appendix~\ref{apx:conjugation} derives the momentum conjugations, and Appendix~\ref{apx:antiunitarity} verifies the antiunitary invariance.

Higher orbitals produce energy shifts and admixture within the symmetry sectors. The accuracy of the four-state description must therefore be assessed through orbital convergence, spectral changes, and the weight retained in $\mathcal P$. The virtual-state calculation in Sec.~\ref{app:mixing} separates these effects from opposite-parity mixing.

\subsection{Coupling between Orbital Shells}\label{app:mixing}
The spatial identity $V_{1,0}=-V_{0,1}$ eliminates the coupling between the two positive-parity Bell states in the four-state projection. The Coulomb interaction nevertheless couples this orbital shell to higher angular momentum states. To evaluate these contributions, we expand $H=H_0+V_{int}$ beyond the four-dimensional unperturbed ground-state manifold.

We first construct the twelve-state space $\mathcal B=\mathcal P\oplus\mathcal Q^+\oplus\mathcal Q^-$. The retained manifold is $\mathcal P=\{|1,1\rangle,|1,2\rangle,|2,1\rangle,|2,2\rangle\}$, with unperturbed energy $E_0=2E_z$. The nearest single-excitation groups are
\begin{align}
\mathcal Q^+&=\{|1,3\rangle,|2,3\rangle,|3,1\rangle,|3,2\rangle\},\label{eq:equation10}\\
\mathcal Q^-&=\{|1,0\rangle,|2,0\rangle,|0,1\rangle,|0,2\rangle\}.\label{eq:Qgroups}
\end{align}
Each state in these groups has kinetic energy $E_z+9E_z=10E_z$, giving the excitation energy $\Delta E_{\mathrm{kin}}=8E_z$. The superscripts distinguish the upper and lower orbital groups; they are separate from the parity labels of the Bell states.

In this ordered basis, the projected interacting Hamiltonian is
\begin{equation}
H_{\mathcal B}=\begin{pmatrix}
H_{\mathcal P}&V_{\mathcal{PQ}^+}&V_{\mathcal{PQ}^-}\\
V_{\mathcal{PQ}^+}^{\dagger}&H_{\mathcal Q^+}&V_{\mathcal Q^+\mathcal Q^-}\\
V_{\mathcal{PQ}^-}^{\dagger}&V_{\mathcal Q^+\mathcal Q^-}^{\dagger}&H_{\mathcal Q^-}
\end{pmatrix}.\label{eq:twelveblock}
\end{equation}
The diagonal blocks include the interactions within their respective orbital groups. The interaction also generally couples $\mathcal Q^+$ and $\mathcal Q^-$. These matrix elements are retained when diagonalizing $H_{\mathcal B}$. The dagger denotes the Hermitian adjoint.

In a perturbative calculation through second order, the virtual resolvent is constructed from $H_0$. An interaction within the virtual space first enters a return path to $\mathcal P$ at third order. This order counting permits its omission from the second-order correction, while leaving it present in the interacting Hamiltonian.

The coupling matrix elements follow from Eq.~\eqref{eq:fourier}: $\langle p_1,p_2|V_{int}|q_1,q_2\rangle=V_{q_1-p_1,q_2-p_2}$. Simultaneous reversal of both transfers gives the same coefficient for the present kernel. Using the stated ordering of $\mathcal P$ and $\mathcal Q^+$ yields
\begin{equation}
V_{\mathcal{PQ}^+}=\begin{pmatrix}
V_{0,2}&V_{1,2}&V_{2,0}&V_{2,1}\\
V_{0,1}&V_{1,1}&V_{2,-1}&V_{2,0}\\
V_{-1,2}&V_{0,2}&V_{1,0}&V_{1,1}\\
V_{-1,1}&V_{0,1}&V_{1,-1}&V_{1,0}
\end{pmatrix}.\label{eq:VPQ}
\end{equation}
The lower coupling block is obtained by the same momentum-transfer prescription. The operation $\mathcal X$ maps the ordered $\mathcal Q^+$ states to the $\mathcal Q^-$ states with indices $(2,1,4,3)$. Appendix~\ref{app:virtual} derives the relation between the two coupling blocks.

The twelve-state space includes the nearest single-excitation groups. Completing the product basis with $p_j\in\{0,1,2,3\}$ also requires $\mathcal Q^d=\{|0,0\rangle,|0,3\rangle,|3,0\rangle,|3,3\rangle\}$. These four double-excitation states have energy $18E_z$ and excitation energy $16E_z=2\Delta E_{\mathrm{kin}}$. Their direct coupling to $\mathcal P$ contributes at second order as well. More distant orbital states provide further virtual contributions.

To obtain the effective low-energy Hamiltonian, let $P$ project onto $\mathcal P$ and let $Q$ project onto the virtual space. For the full problem, $Q=1-P$; a finite calculation uses the corresponding retained virtual projector. We choose an anti-Hermitian generator $\mathcal S^\dagger=-\mathcal S$, with $P\mathcal SP=Q\mathcal SQ=0$, and define $\widetilde H_{\mathcal P}=P e^{-\mathcal S}H e^{\mathcal S}P$. Writing $V_{\mathrm{off}}=PV_{int}Q+QV_{int}P$, the first-order cancellation condition is $[H_0,\mathcal S]=-V_{\mathrm{off}}$.

For $|a\rangle\in\mathcal P$ and $|d\rangle\in Q$, this condition gives $\mathcal S_{ad}=-V_{ad}/(E_a^{(0)}-E_d^{(0)})$, where $V_{ad}=\langle a|V_{int}|d\rangle$ and the superscript $(0)$ denotes an unperturbed kinetic energy. Expansion of the transformed Hamiltonian gives $\delta H^{(2)}=\tfrac12P[V_{\mathrm{off}},\mathcal S]P$. Thus
\begin{equation}
\begin{aligned}
\delta H_{ab}^{(2)}={}&\frac12\sum_{d\in Q}V_{ad}V_{db}\\
&\times\left[\frac1{E_a^{(0)}-E_d^{(0)}}+\frac1{E_b^{(0)}-E_d^{(0)}}\right].
\end{aligned}
\label{eq:SWmatrix_main}
\end{equation}
The retained unperturbed states are degenerate, $E_a^{(0)}=E_b^{(0)}=E_0$. Equation~\eqref{eq:SWmatrix_main} therefore reduces to the second-order operator
\begin{equation}
\delta H^{(2)}=-PV_{int}Q\bigl[Q(H_0-E_0)Q\bigr]^{-1}QV_{int}P,\label{eq:SW}
\end{equation}
where the inverse acts within the virtual space. These are the unperturbed denominators of the interaction expansion. Appendix~\ref{apx:sw_generator} derives the second-order matrix elements from the generator.

For the nearest single-excitation groups, the denominators are exactly $\Delta E_{\mathrm{kin}}$, so their contribution is
\begin{equation}
\delta H_{\mathrm{single}}^{(2)}=-\frac{V_{\mathcal{PQ}^+}V_{\mathcal{PQ}^+}^{\dagger}+V_{\mathcal{PQ}^-}V_{\mathcal{PQ}^-}^{\dagger}}{\Delta E_{\mathrm{kin}}}.\label{eq:SWsingle_main}
\end{equation}
The double-excitation contribution is $-V_{\mathcal{PQ}^d}V_{\mathcal{PQ}^d}^{\dagger}/(2\Delta E_{\mathrm{kin}})$. Together, these terms give the second-order correction within the sixteen-state product-space truncation. Higher virtual shells retain their individual kinetic denominators.

The Hermitian products in these expressions preserve the relative signs of the transition amplitudes. For example, $(V_{\mathcal{PQ}^+}V_{\mathcal{PQ}^+}^{\dagger})_{ab}=\sum_{d\in\mathcal Q^+}V_{ad}V_{bd}^*$. Only a diagonal entry is necessarily a sum of absolute squares; an off-diagonal entry contains products between different transition amplitudes.

To establish the parity constraint, let $|a,+\rangle$ and $|b,-\rangle$ be retained parity eigenstates, and pair each virtual orbital state $|d\rangle$ with $|\bar d\rangle=\mathcal X|d\rangle$. Since $\mathcal X$ commutes with both $H_0$ and $V_{int}$, $E_{\bar d}^{(0)}=E_d^{(0)}$, $\langle a,+|V_{int}|\bar d\rangle=\langle a,+|V_{int}|d\rangle$, and $\langle\bar d|V_{int}|b,-\rangle=-\langle d|V_{int}|b,-\rangle$. The two products therefore cancel with equal denominators.

For a virtual space invariant under $\mathcal X$, the complete correction satisfies
\begin{equation}
[\delta H^{(2)},\Sigma_x]=0,\qquad
\langle\Psi_i^+|\delta H^{(2)}|\Psi_j^-\rangle=0.\label{eq:SWzero}
\end{equation}
The contributions of the two nearest single-excitation groups cancel in every opposite-parity matrix element. The double-excitation group is itself invariant under $\mathcal X$. Appendix~\ref{apx:sw_pairs} gives the resulting Bell-basis correction. Table~\ref{tab:SW} shows the opposite signs of the two single-excitation contributions for one representative matrix element. The cancellation extends beyond second order: since $P$ and $Q$ commute with $\mathcal X$, the exact energy-dependent effective Hamiltonian commutes with $\Sigma_x$ wherever its virtual-space resolvent exists (Appendix~\ref{apx:sw_resolvent}). Opposite-parity mixing is therefore excluded in the symmetric model, while energy shifts and higher-orbital admixture within each sector remain allowed.

To characterize selected higher-transfer amplitudes, we retain the dimensionless diagnostic
\begin{equation}
\begin{aligned}
u_{\mathrm{cross}}(\alpha,\beta;\rho_{\mathrm{soft}})
&=\max_{(m,n)\in\{(0,2),(1,2),(-1,2)\}}\left|\frac{V_{m,n}}{E_c}\right|.
\end{aligned}\label{eq:ucross_selected}
\end{equation}
Here the coefficients are evaluated using the softened kernel with the indicated $\rho_{\mathrm{soft}}$. This diagnostic measures the three specified transfer channels; the full coupling matrix contains all retained transfers.

\begin{figure}[tbp]\centering
\includegraphics[width=\columnwidth]{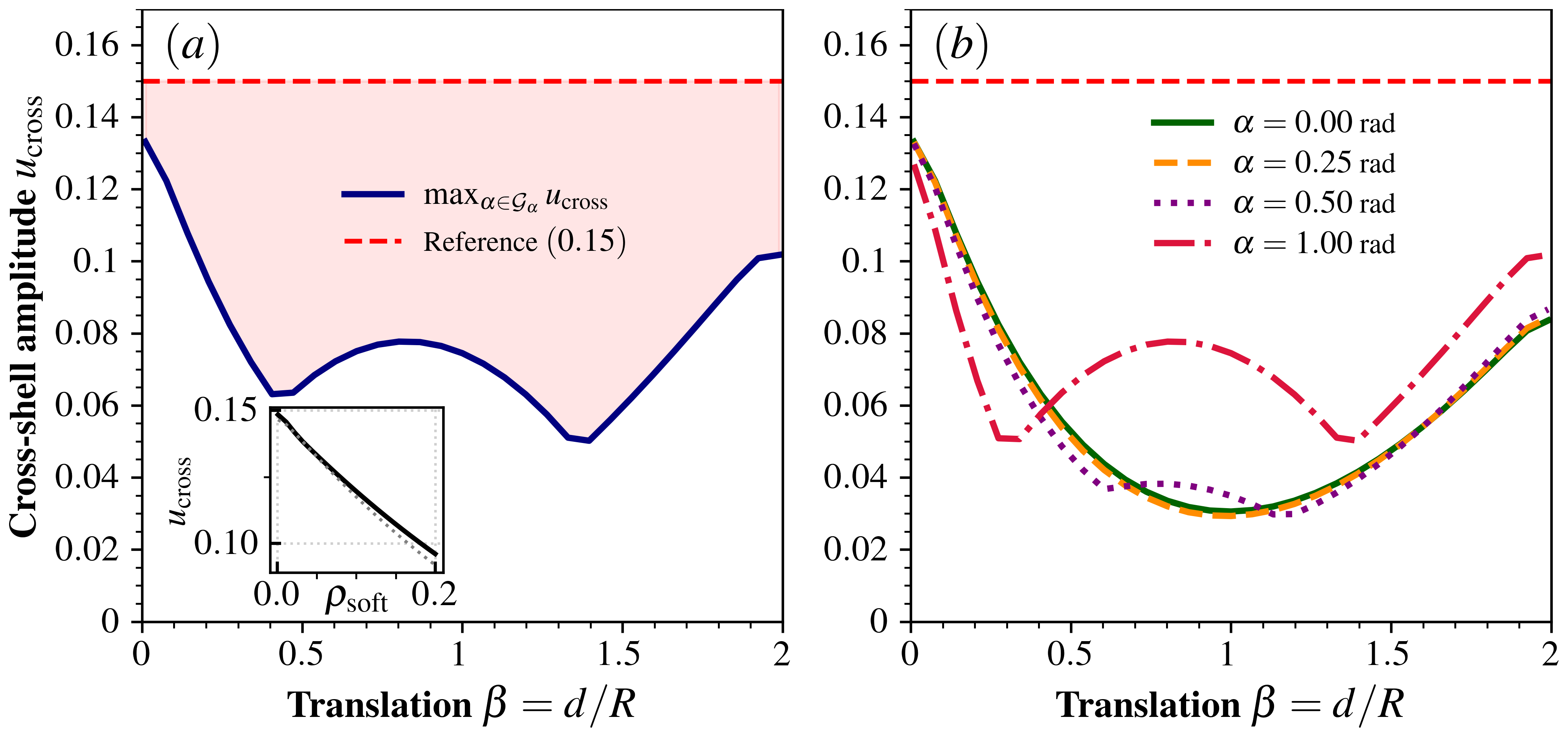}
\caption{Selected higher-orbital Fourier amplitudes for the softened interaction with $\rho_{\mathrm{soft}}=0.05$. At each geometry, $u_{\mathrm{cross}}$ is the largest coefficient magnitude in units of $E_c$ for transfers $(0,2)$, $(1,2)$, and $(-1,2)$. (a) Maximum over 31 uniformly spaced tilts $0\leq\alpha\leq1$ rad, evaluated at 31 uniformly spaced translations $0.01\leq\beta\leq1.99$. The inset shows the maximum over a $61\times61$ grid on the same geometric domain for 20 uniformly spaced softening parameters $0.001\leq\rho_{\mathrm{soft}}\leq0.2$ (solid), with the fixed geometry $(\alpha,\beta)=(0.5,0.01)$ retained as a dotted reference. (b) Separately evaluated slices at $\alpha=0,0.25,0.50,1.00$ rad. The dashed line at $0.15$ is a reference value for this sampled calculation. Refined periodic quadrature and selected independent adaptive checks are specified in Appendix~\ref{app:numerics}.}
\label{fig:mixing}\end{figure}
Figure~\ref{fig:mixing}(a) shows the maximum of the three-channel diagnostic in Eq.~\eqref{eq:ucross_selected} over the 31 sampled tilts at each translation; panel (b) shows slices at fixed tilt. The inset follows the sampled geometric maximum as softening varies; the dotted curve gives the fixed-geometry reference. Nonzero values of these coefficients quantify inter-shell coupling and are compatible with the opposite-parity cancellation in Eq.~\eqref{eq:SWzero}.

To bound the second-order correction, we use the entire coupling operator between the retained and virtual spaces. With $\|\cdot\|$ denoting the induced operator norm and every virtual excitation at least $\Delta E_{\mathrm{kin}}$ above $E_0$, Eq.~\eqref{eq:SW} gives
\begin{eqnarray}
\|\delta H^{(2)}\|&\leq&\frac{\|PV_{int}Q\|^2}{\Delta E_{\mathrm{kin}}} \nonumber \\
&=&\lambda E_c\left\|\frac{PV_{int}Q}{E_c}\right\|^2.\label{eq:SWnorm}
\end{eqnarray}
The bound includes parity-preserving energy shifts.

At fixed disjoint geometry, the unsoftened multiplication operator is bounded, with $\|V_{int}\|=E_c/\min_{\theta_1,\theta_2}\mathcal D$. A conservative weak-coupling condition is therefore $\|V_{int}\|/\Delta E_{\mathrm{kin}}\ll1$. Beyond a perturbative treatment, enlarged-basis diagonalization and orbital convergence checks quantify the accuracy of the low-energy description. The selected Fourier amplitudes characterize orbital coupling, while the internal Bell splitting describes the separation of two symmetry-selected branches.

\section{Geometric Tilt and Orbital Concurrence}\label{sec:concurrence}
\begin{figure}[tbp]\centering
\includegraphics[width=\columnwidth]{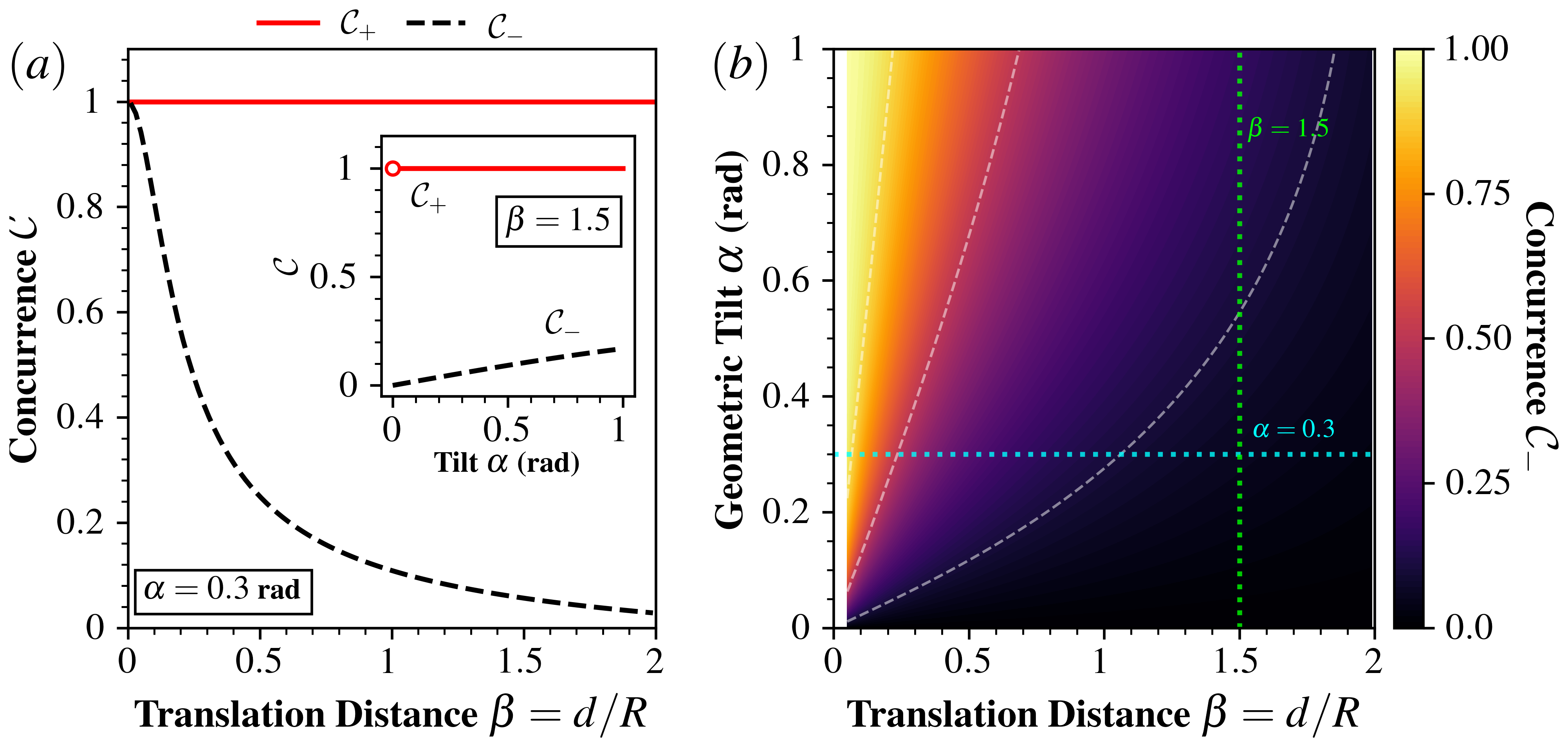}
\caption{Concurrence of the four-state eigenvectors for the unsoftened interaction ($\rho_{\mathrm{soft}}=0$). (a) $\mathcal C_\pm$ at $\alpha=0.3$ rad, evaluated on 101 uniformly spaced translations $0.01\leq\beta\leq1.99$. The inset uses $\beta=1.5$ and 101 uniformly spaced tilts $0\leq\alpha\leq1$ rad. The open point at $(\alpha,\mathcal C_+)=(0,1)$ denotes the limiting Bell branch; exact zero tilt leaves the positive-parity block degenerate. (b) $\mathcal C_-$ on a $101\times101$ grid spanning $0\leq\alpha\leq1$ rad and $0.05\leq\beta\leq1.99$. The unfilled strip at smaller $\beta$ lies outside the sampled grid. Cyan and green lines mark the parameter slices used in panel (a). Angular quadrature is refined using the coefficient and concurrence criteria in Appendix~\ref{app:numerics}.}
\label{fig:concurrence}\end{figure}

The concurrence expressions derived in Eqs.~\eqref{eq:Cplus_parity} and \eqref{eq:Cminus_parity} quantify the orbital entanglement of the four-state eigenvectors. For $W\neq0$, the separate positive-parity eigenstates remain Bell states with $\mathcal C_+=1$. The negative-parity concurrence depends on the relative magnitudes of $W$ and $V_{1,0}$.

Figure~\ref{fig:concurrence}(a) shows translation and tilt scans; panel (b) maps the negative-parity concurrence over both parameters. The negative-parity concurrence is reduced when $|V_{1,0}|$ is large relative to $|W|$ and increases as their ratio decreases. The curves describe the geometry-dependent entanglement of the projected eigenstates.

Within this projection, the common Coulomb energy scale multiplies the entire interaction matrix, while $H_0=E_0\mathbb I$. Consequently, the eigenvectors at nonzero interaction depend on the geometry and coefficient ratios and are invariant under a common rescaling of the interaction. The parameter $\lambda$ becomes relevant to their higher-orbital dressing and to the accuracy of the four-state approximation.

At exact orthogonality, $\alpha=0$, the geometric exchange difference vanishes and $H^+$ becomes degenerate. This removes the selection of individual positive-parity eigenvectors by the projected Hamiltonian. The corresponding concurrence depends on the state prepared within that eigenspace.

\subsection{Concurrence at Exact Orthogonality}\label{app:orthogonality}
At $\alpha=0$, the dimensionless distance reduces to $\mathcal D_s=\sqrt{\mathcal D_s^2(\theta_1,\theta_2;\beta)}$. The geometric exchange energy is
\begin{equation}
\begin{aligned}
W(0,\beta)={}&\frac{E_c}{2(2\pi)^2}\int_{-\pi}^{\pi}\!d\theta_1\int_{-\pi}^{\pi}\!d\theta_2\\
&\times\frac{e^{i(\theta_1+\theta_2)}-e^{i(\theta_1-\theta_2)}}{\mathcal D_s(\theta_1,\theta_2;\beta)}.
\end{aligned}\label{eq:Wzero}
\end{equation}
The numerator separates into real and imaginary parts as $e^{i(\theta_1+\theta_2)}-e^{i(\theta_1-\theta_2)}=-2\sin\theta_1\sin\theta_2+2i\cos\theta_1\sin\theta_2$.

The denominator depends only on $\cos\theta_1$ and $\cos\theta_2$ and is even under each angular inversion independently. The imaginary contribution is odd in $\theta_2$ and integrates to zero. The real contribution is also odd in $\theta_2$ for fixed $\theta_1$, and its integral vanishes. Thus $V_{1,1}(0,\beta)=V_{1,-1}(0,\beta)$ and $W(0,\beta)=0$.

At zero tilt the single-ring reflections $\mathcal X_j|p_j\rangle=|3-p_j\rangle$ also commute with the full Hamiltonian: each preserves its shifted kinetic energy and inverts one angle of the even kernel. Since $\mathcal X=\mathcal X_1\mathcal X_2$ and $\mathcal Y\mathcal X_1\mathcal Y=-\mathcal X_2$, the operation $\mathcal X_1$ exchanges the two $\mathcal Y$ sectors at the same energy within $\mathcal X=+1$.

In the four-state space, $V_{1,0}+V_{0,1}=0$ then gives $H^+=(V_{0,0}+V_{1,1})\mathbb I$. Every normalized vector in the positive-parity two-dimensional space is then an eigenvector. The Bell state $|\Psi_1^+\rangle$ has concurrence one, whereas
\begin{equation}
\begin{aligned}
\frac{|\Psi_1^+\rangle+|\Psi_2^+\rangle}{\sqrt2}
={}&\left(\frac{|1\rangle+|2\rangle}{\sqrt2}\right)_1\\
&\otimes\left(\frac{|1\rangle+|2\rangle}{\sqrt2}\right)_2
\end{aligned}\label{eq:zeroexample}
\end{equation}
is separable. The degenerate eigenspace therefore contains states with different concurrences, selected by preparation or by an additional perturbation.

A specified preparation or perturbation determines the state selected at degeneracy. Along the separate Bell eigenbranches approaching zero tilt, the concurrence remains one. The open point in the inset of Fig.~\ref{fig:concurrence} distinguishes this branch limit from an unspecified state at $\alpha=0$.

For the negative-parity block at $\alpha=0$, Eq.~\eqref{eq:Cminus_parity} gives $\mathcal C_-=0$ whenever $V_{1,0}\neq0$. If $V_{1,0}$ also vanishes, that block is degenerate and its eigenstate must likewise be specified. These conclusions distinguish the closing of an internal spectral gap from the entanglement of a chosen state.

\subsection{Entanglement in the enlarged orbital space}\label{sec:fullentropy}
For $|\psi\rangle=\sum_{p,q}c_{pq}|p,q\rangle$, with normalized orbital amplitudes $c_{pq}$, let $C=(c_{pq})$. The full ring-mode entropy is:
\begin{equation}
 \varrho_1=CC^\dagger,\qquad S=-\operatorname{Tr}(\varrho_1\log_2\varrho_1).\label{eq:entropy}
\end{equation}
Here $\operatorname{Tr}$ denotes the trace, $\varrho_1$ is the reduced density operator of ring 1 and $S$ is measured in bits. This pure-state entanglement measure applies to the enlarged local orbital spaces~\cite{NielsenChuang2000}. For the four-state encoding we use
\begin{align}
 w_{\mathcal P}&=\langle\psi|P|\psi\rangle,\qquad
 |\psi_{\mathcal P}\rangle=P|\psi\rangle/\sqrt{w_{\mathcal P}},\label{eq:weight}\\
 \mathcal C_{\mathcal P}&=\frac{2|c_{11}c_{22}-c_{12}c_{21}|}{w_{\mathcal P}},
 \qquad w_{\mathcal P}>0.\label{eq:conditional}
\end{align}
The latter is the concurrence of the normalized two-qubit projection (Appendix~\ref{apx:partial_trace}). For this pure state, unit concurrence is equivalent to one bit of entanglement. The full entropy measures entanglement between the two rings with all retained orbitals included, while $w_{\mathcal P}$ gives the probability of occupying the four-state space.

\begin{figure}[tbp]\centering
\includegraphics[width=\columnwidth]{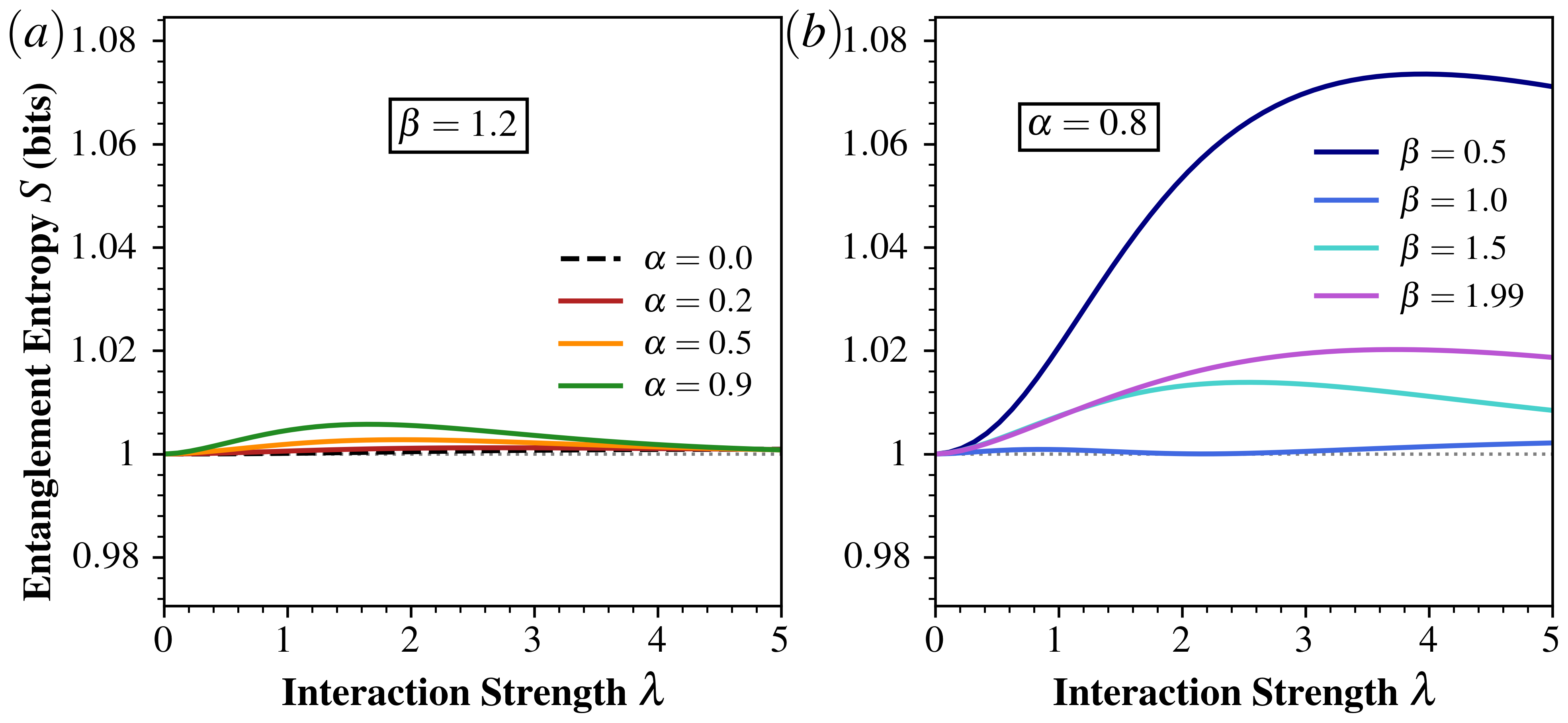}
\caption{Full ring-mode entropy along the branch connected to $\Psi_2^+$. At each point the state is the lowest eigenstate in the joint sector $(\mathcal X,\mathcal Y)=(+1,-1)$, with $\rho_{\mathrm{soft}}=0$ and 50 uniformly spaced values $0.01\leq\lambda\leq5$. (a) $\alpha=0,0.2,0.5,0.9$ rad at $\beta=1.2$. The zero-tilt curve uses this joint-symmetry selection within the degenerate space. (b) $\beta=0.5,1.0,1.5,1.99$ at $\alpha=0.8$ rad. The dotted line marks the exact one-bit lower bound in this sector. The orbital window is $p=1-K,\ldots,2+K$ with $K=10$; cutoff and quadrature checks are described in Appendix~\ref{app:numerics}.}
\label{fig:fullentropy}\end{figure}
The state tracked in Fig.~\ref{fig:fullentropy} belongs to $(\mathcal X,\mathcal Y)=(+1,-1)$. The $\mathcal Y=-1$ condition makes the orbital coefficient matrix antisymmetric up to a local unitary transformation. In a finite symmetry-preserving orbital window, Youla's normal form implies that the nonzero singular values occur in pairs~\cite{Youla1961}. Their squares are the Schmidt weights, so $S\geq1$ bit. Appendix~\ref{app:observable_steps} extends this argument to the full orbital space. This lower bound applies to every normalized pure state in that symmetry sector, including a global ground state when it belongs to the sector. States outside the sector require a separate entanglement analysis. Within $\mathcal P$ the joint sector is one dimensional, giving $\mathcal C_{\mathcal P}=1$ for nonzero projection. Table~\ref{tab:cutoff} compares orbital cutoffs at a separate reference geometry. Across the sampled parameters, the entropy reaches about $1.07$ bits, while the four-state weight falls to about $0.57$. In Fig.~\ref{fig:fullentropy}(b), the $\beta=1$ branch remains close to one bit, reaching $S\simeq1.0021$ bits at $\lambda=5$, where $w_{\mathcal P}\simeq0.5896$. Thus a small change in entropy can accompany a substantial redistribution among orbitals. In this sector $P|\psi\rangle=a|\Psi_2^+\rangle$ with complex amplitude $a$, so $|\langle\Psi_2^+|\psi\rangle|^2=|a|^2=w_{\mathcal P}$. An ideal local projection onto the two retained orbitals of the selected spin branch, $\operatorname{span}\{|1\rangle,|2\rangle\}$ on each ring, preserving coherence within each retained subspace, therefore succeeds jointly with probability $w_{\mathcal P}$. Conditional on success, the normalized state is $|\Psi_2^+\rangle$ up to an overall phase. Orbital admixture can therefore lower the probability of obtaining this Bell pair while the full entropy still satisfies its lower bound. The numerical calculation characterizes stationary eigenstate entanglement. The same lower bound also holds throughout closed-system evolution from a pure $\mathcal Y=-1$ state whenever the full Hamiltonian preserves $\mathcal Y$ at every time; Appendix~\ref{apx:entropy_bound} gives the propagator argument. The entropy and encoding weight can vary during this evolution. A coherence lifetime requires a specified dynamical noise model.

\subsection{Relative localization and joint angular correlations}\label{sec:localization}
To characterize the global ground state independently of the selected Bell branch, we use the relative circular moment, motivated by position-phase observables in periodic systems~\cite{Resta1998,Resta2006},
\begin{equation}
 \begin{aligned}Z_{\mathrm{rel}}&=\langle e^{i(\theta_1-\theta_2)}\rangle
 =\sum_{p,q}c_{p+1,q-1}^*c_{pq},\\
 \lambda_{rel}^2&=-\ln|Z_{\mathrm{rel}}|^2.\end{aligned}\label{eq:spread}
\end{equation}
$\lambda_{rel}^2$ is dimensionless and distinct from the coupling $\lambda$. It measures the first circular harmonic of the relative angular distribution and diverges when $Z_{\mathrm{rel}}=0$.

\begin{figure}[tbp]\centering
\includegraphics[width=\columnwidth]{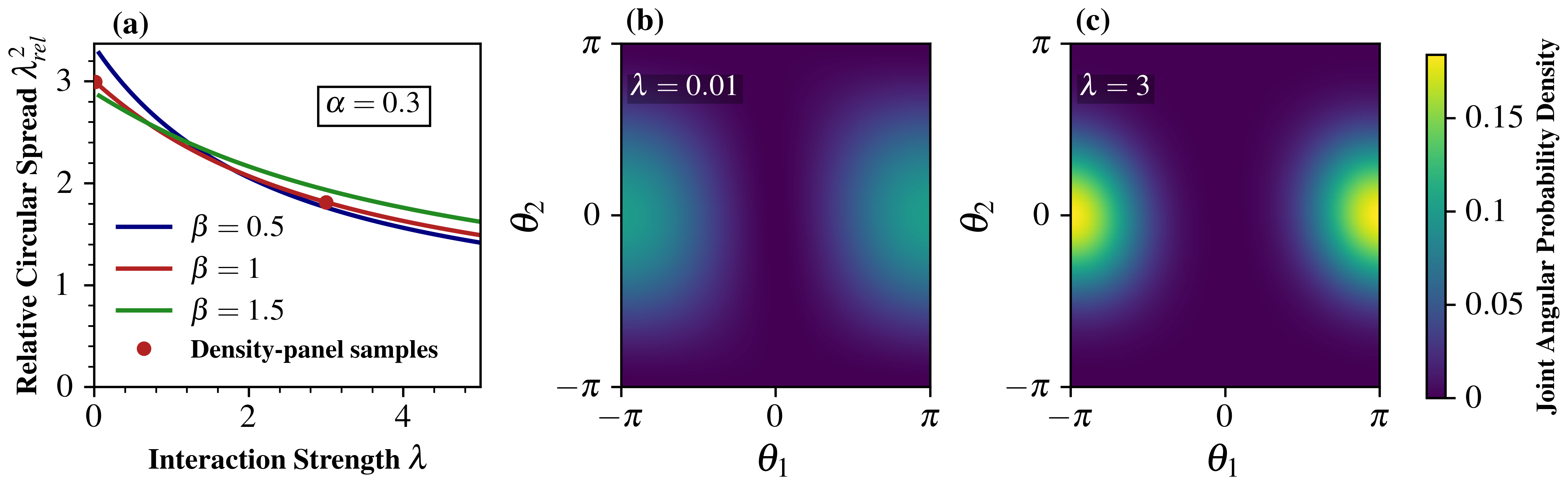}
\caption{Global-ground-state angular correlations at
$\alpha=0.3$ rad and $\rho_{\mathrm{soft}}=0$.
(a) Relative circular spread for $\beta=0.5,1.0,1.5$,
evaluated at the 79 positive interaction strengths
$\lambda=5j/79$, $j=1,\ldots,79$.
(b,c) Joint angular densities at $\beta=1$ and
$\lambda=0.01$ and $3$, respectively, with a shared color
scale. Red circles in panel (a), labelled
``Density-panel samples,'' show the separately computed
spreads for the ground states displayed in panels (b)
and (c).
The densities are normalized with respect to
$d\theta_1\,d\theta_2$ and reconstructed on a
$256\times256$ cell-centered grid. Both density calculations
satisfy the convergence criteria for density, energy,
gap, and spread upon increasing the orbital cutoff
from $K=8$ to $K=10$. Quadrature and cutoff criteria
are specified in Appendix~\ref{app:numerics}.}
\label{fig:localization}\end{figure}
The coordinate wavefunction and joint angular density are:
\begin{equation}
 \psi(\theta_1,\theta_2)=\frac1{2\pi}\sum_{p,q}c_{pq}e^{i(p\theta_1+q\theta_2)},
 \qquad \mathcal P_{ang}=|\psi|^2.\label{eq:density}
\end{equation}
The integral of $\mathcal P_{ang}$ is one, and its relative angular moment gives Eq.~\eqref{eq:spread}. Appendix~\ref{apx:angle_moment} derives this relation from Fourier orthogonality. Figure~\ref{fig:localization}(a) shows the decrease in circular spread; panels (b) and (c) show the corresponding angular densities at weak and strong coupling. These observables describe the angular distribution of the two-electron ground state. The curves begin at positive coupling because the kinetic ground state is degenerate at $\lambda=0$.

\subsection{Sensitivity to a specified spatial perturbation}\label{sec:robustness}
We examine an explicit angular potential on ring 1,
$V_{dis}=\epsilon E_c\cos\theta_1$.
Here $\epsilon$ is a dimensionless potential amplitude, separate from $\epsilon_{\mathrm{eff}}$. This deterministic perturbation preserves $\mathcal X$ and breaks $\mathcal Y$. Within $\mathcal P$ it adds $\epsilon E_c/2$ to the off-diagonal entry of $H^+$. The four-state result is consequently
\begin{equation}
 \mathcal C_{+,\mathcal P}^{(4)}
 =\frac{|W|}{\sqrt{W^2+(\epsilon E_c/2)^2}}.\label{eq:Cpert}
\end{equation}
It explicitly selects separable positive-block eigenvectors at $W=0$ and $\epsilon\ne0$. Taking $\epsilon\to0$ first at $W\ne0$ selects the Bell limit, displaying the importance of the perturbation and limiting prescription.

\begin{figure}[tbp]\centering
\includegraphics[width=\columnwidth]{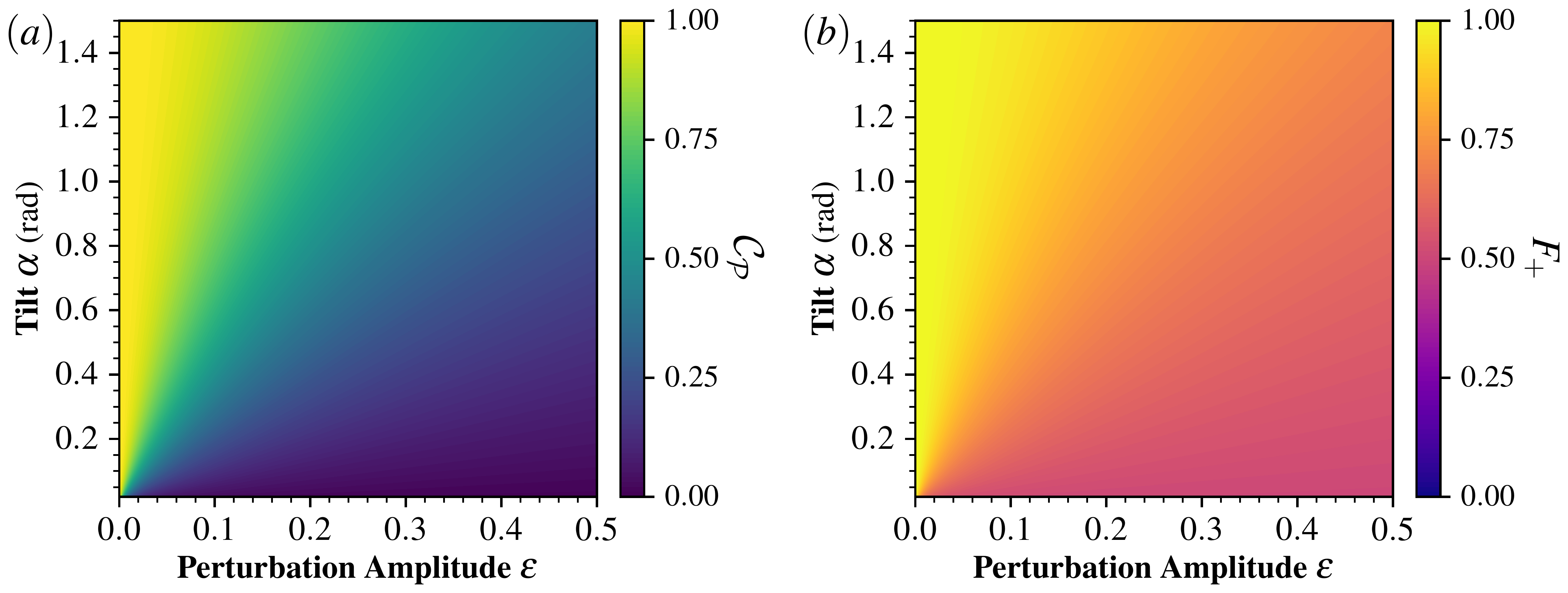}
\caption{Response to $V_{dis}=\epsilon E_c\cos\theta_1$ for $\beta=1.1$, $\lambda=0.25322$, and $\rho_{\mathrm{soft}}=0$. The grid uses 250 uniformly spaced tilts $0.02\leq\alpha\leq1.5$ rad and 250 amplitudes $0\leq\epsilon\leq0.5$. The state is the lowest nondegenerate eigenstate in the retained $\mathcal X=+1$ sector. (a) Concurrence of its normalized four-state projection. (b) Full orbital fidelity $F_+=|\langle\psi_+(0,\alpha)|\psi_+(\epsilon,\alpha)\rangle|^2$ relative to the unperturbed state at the same tilt. Orbital windows are refined through $K=10,12,16,20$ as required by the convergence tests; the projection weight remains above $0.98$. The perturbation is a deterministic one-body angular potential on the circular-ring geometry.}
\label{fig:robustness}\end{figure}
Figure~\ref{fig:robustness}(a) shows a continuous reduction of conditional concurrence as the one-body perturbation becomes comparable to the internal splitting. In the most affected region the concurrence approaches zero and the reference-state fidelity in panel (b) approaches one half, while the four-state weight remains above $0.98$. The potential preserves the circular geometry and its Coulomb identity $V_{1,0}+V_{0,1}=0$, while breaking the additional $\mathcal Y$ symmetry of the full Hamiltonian. The admixture depends on the perturbation matrix elements and the energy gaps to the states they couple.

For a perturbation $\eta A$ within the domain of first-order perturbation theory, with dimensionless real amplitude $\eta$ and Hermitian energy operator $A$, the parity-odd part is $A_o=(A-\mathcal X A\mathcal X)/2$. Away from degeneracies, the first-order opposite-parity component of a state $|a,+\rangle$ is
\begin{equation}
 |\delta\psi_a^-\rangle=\eta\sum_{b,-}
 \frac{|b,-\rangle\langle b,-|A_o|a,+\rangle}{E_{a,+}-E_{b,-}}.\label{eq:paritypert}
\end{equation}
Weak opposite-parity admixture requires the squared norm of Eq.~\eqref{eq:paritypert} to be much smaller than unity. Its breakdown depends on the actual energy differences between the coupled sectors and the specified perturbation. Ring deformation, angular disorder, and flux detuning act through different matrix elements. Appendix~\ref{app:imperfections} evaluates the leading Coulomb-kernel correction for an elliptic deformation of ring 1 and gives the general shape derivative and a flux-detuning term. The shape calculation isolates the Coulomb-kernel variation at fixed angular basis and measure. A complete deformed-ring treatment additionally requires the modified kinetic operator and orbital basis. Degenerate levels require diagonalization in the coupled subspace. Equal flux detunings on the two rings preserve $\mathcal Y$ even though they break the original $\mathcal X$ symmetry. The one-bit bound therefore remains valid for pure states in the $\mathcal Y=-1$ sector under this perturbation, whereas unequal detunings generally break that symmetry as well (Appendix~\ref{app:imperfections}).

\section{Physical Implementation and Interaction Scales}\label{sec:implementation}
\begin{figure}[tbp]\centering
\includegraphics[width=\columnwidth]{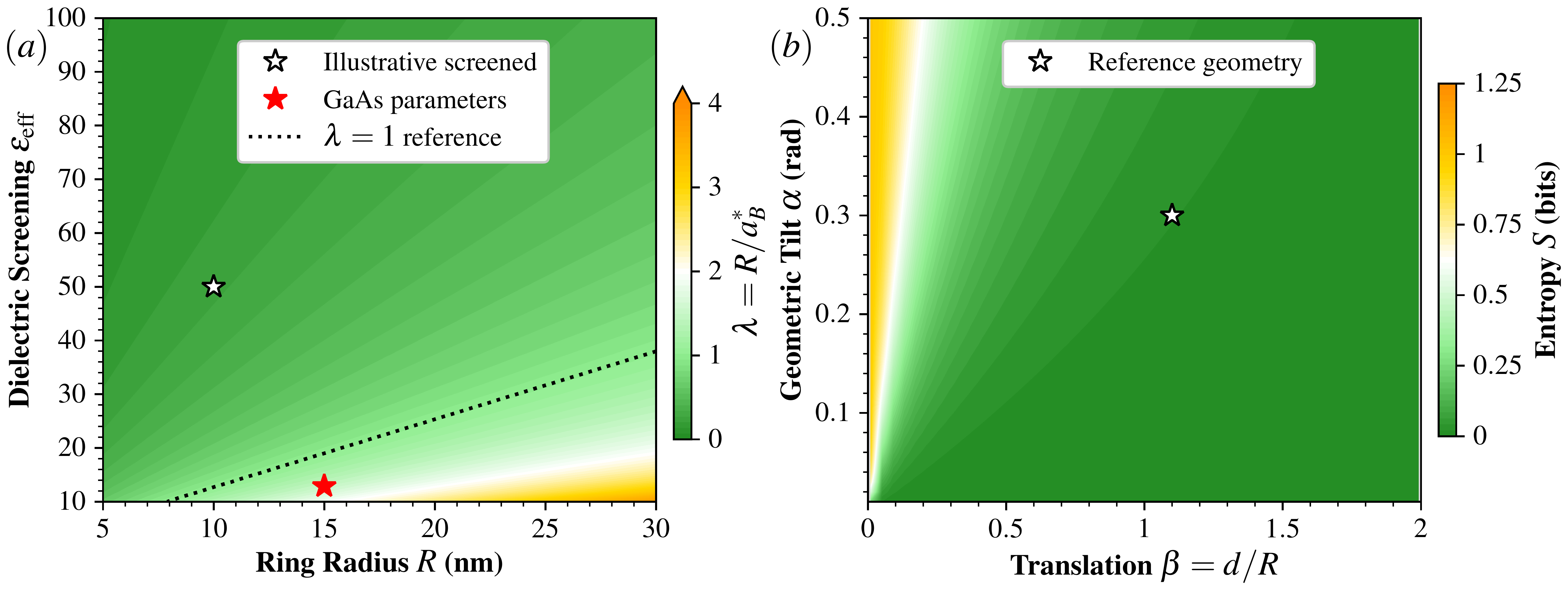}
\caption{Interaction scale and ground-state orbital entanglement. (a) $\lambda=R/a_B^*$ for $m^*=0.067m_e$, evaluated on a $50\times50$ grid spanning $5\leq R\leq30$ nm and $10\leq\epsilon_{\mathrm{eff}}\leq100$. The dotted line marks the reference value $\lambda=1$. Stars indicate $(R,\epsilon_{\mathrm{eff}})=(10\ \mathrm{nm},50)$ and $(15\ \mathrm{nm},12.9)$. (b) Ground-state ring-mode entropy at $\lambda=0.25322$, corresponding to the former parameter pair. The $50\times50$ geometric grid spans $0.01\leq\alpha\leq0.5$ rad and $0.01\leq\beta\leq1.99$, with $\rho_{\mathrm{soft}}=0$. The white star marks $(\beta,\alpha)=(1.1,0.3)$. Separate colorbars show the interaction ratio and entropy in bits. Orbital and angular convergence criteria are specified in Appendix~\ref{app:numerics}.}
\label{fig:phasediagram}\end{figure}

To translate the dimensionless parameters $(\lambda,\beta,\alpha)$ into physical scales, we consider a GaAs-like effective mass $m^*=0.067m_e$, where $m_e$ is the free-electron mass. For a homogeneous dielectric environment, the interaction parameter can be expressed as
\begin{equation}
\lambda=\frac{m^*}{m_e}\frac{R}{\epsilon_{\mathrm{eff}}a_0},
\qquad a_0\simeq0.052918\ \mathrm{nm},\label{eq:materiallambda}
\end{equation}
where $a_0$ is the vacuum Bohr radius. This follows from $\lambda=R/a_B^*$ and $a_B^*=\epsilon_{\mathrm{eff}}(m_e/m^*)a_0$.

For $R=15$ nm and $\epsilon_{\mathrm{eff}}=12.9$, the interaction parameter is $\lambda\simeq1.47$. The overall Coulomb scale is then comparable to the kinetic excitation scale. A weak-coupling SW approximation therefore requires additional justification from the relevant matrix elements and spectral separations. The exact parity symmetry of the ideal orbital Hamiltonian remains independent of this perturbative approximation.

For the illustrative parameters $R=10$ nm and $\epsilon_{\mathrm{eff}}=50$, the interaction ratio decreases to $\lambda\simeq0.2532$. Even at this weaker coupling, the accuracy of the four-state description depends on geometry and higher-orbital admixture. In particular, a small minimum separation can enhance interaction matrix elements. The reference $\lambda=1$ in Fig.~\ref{fig:phasediagram}(a) marks equality of the two defining energy scales.

Figure~\ref{fig:phasediagram}(b) shows the ring-mode entropy of the global ground state, evaluated from Eq.~\eqref{eq:entropy}. At every sampled point this ground state is nondegenerate, has $\mathcal X=-1$, and has nonzero weight in $\mathcal P$. Its projection therefore lies in the negative-parity Bell block, where $\mathcal Y=+1$. Since $P$ and $H$ commute with $\mathcal Y$, the full ground state also has $\mathcal Y=+1$. The one-bit bound applies to the $\mathcal Y=-1$ sector illustrated in Fig.~\ref{fig:fullentropy}.

The dielectric constant in Eq.~\eqref{eq:materiallambda} is a model input. A nearby metallic gate modifies the electrostatic Green function according to its position and shape. Its effect on the noncoplanar rings generally requires a spatially resolved interaction kernel. The value $\epsilon_{\mathrm{eff}}=50$ is used here as an illustrative homogeneous screening parameter. Any proposed screening arrangement must also be checked against the geometric symmetries used in the orbital analysis.

The gauge construction in Sec.~\ref{app:gauge} requires $\lambda_R=\sqrt3\hbar^2/(2m^*R)$. At $R=10$ nm and $m^*=0.067m_e$, this gives $\lambda_R\simeq9.85\times10^{-11}\ \mathrm{eV\,m}$. Rashba coupling and spin-geometric phases provide relevant physical ingredients~\cite{Nagasawa2013,Bihlmayer2022}. An implementation must combine this coupling with an aligned exchange texture, linked confinement, and negligible inter-ring tunneling.

Accurate spin projection requires the couplings between the selected and discarded branches to be small compared with the energy separations of the levels they connect. Moreover, the conical texture has a nonzero average axial component. Its magnetic environment and associated perturbations must therefore be evaluated explicitly. A device calculation must treat confinement, electrostatics, and the spin-dependent terms self-consistently.

\subsection{Evanescent Wavefunction Tunneling and Exchange}\label{app:tunneling}
The ring-mode description of Sec.~\ref{sec:bipartition} assumes one electron in each ring and negligible transfer between the two confinements. A finite material separation can produce evanescent wavefunction overlap and additional exchange interactions. Their magnitudes depend on the transverse confinement and barrier structure~\cite{Burkard1999,Pedersen2007}.

The shortest material gap $d_{\mathrm{gap}}$ is determined by the centerline clearance and the transverse dimensions. For equal circular transverse radii $a_\perp$ in a nonoverlapping tubular model, the gap is $d_{\mathrm{gap}}=R\min_{\theta_1,\theta_2}\mathcal D-2a_\perp$. The tunneling distance therefore depends jointly on translation, tilt and transverse confinement.

For a uniform barrier whose potential energy $V_b$ exceeds the occupied transverse level $E$, the evanescent decay constant is $\kappa=\sqrt{2m_b^*(V_b-E)}/\hbar$, where $m_b^*$ is the barrier effective mass. The overlap has the approximate exponential dependence $\mathcal S_{\mathrm{ov}}\sim\exp(-\kappa d_{\mathrm{gap}})$. Here $\mathcal S_{\mathrm{ov}}$ denotes the spatial overlap, separately from the entropy $S$ and the SW generator $\mathcal S$. For a nonuniform barrier, the decay exponent instead involves the integral of the local decay constant along the tunneling path.

A schematic overlap-squared estimate, $J_{\mathrm{direct}}\sim E_c|\mathcal S_{\mathrm{ov}}|^2$, illustrates an exponential exchange scale. Its prefactor depends on the wavefunctions and Coulomb integrals. The total exchange coupling also includes tunneling-mediated contributions that depend on tunneling amplitudes and charging energies.

For an illustrative barrier $V_b-E=4$ eV with $m_b^*=m_e$, the decay constant is $\kappa\simeq10.25\ \mathrm{nm}^{-1}$. At $d_{\mathrm{gap}}=1$ nm, the corresponding overlap-squared exponential factor is $\exp(-2\kappa d_{\mathrm{gap}})\simeq1.3\times10^{-9}$. This example demonstrates the sensitivity to barrier height and gap width. It assumes those barrier parameters and omits device-dependent prefactors. A finite barrier leaves a finite overlap; a hard-wall confinement is a separate idealization.

Any residual tunneling or exchange must be compared with the states it connects and their energy separations. The internal Bell splitting $\Delta E_+=2|W|$ is the relevant denominator only for a perturbation coupling the corresponding Bell levels. Its value can become small and vanishes at exact orthogonality in the four-state model (Sec.~\ref{app:orthogonality}). This comparison requires the geometry-dependent value of $W$.

Finite tunneling can also admix the occupation sectors $(2,0)$ and $(0,2)$ into the one-electron-per-ring sector $(1,1)$. Let $\Delta_{\mathrm{ch}}>0$ denote the minimum energy separation from the relevant $(1,1)$ states to accessible charge-transfer states in the absence of tunneling. This separation depends on charging energies, inter-ring repulsion, orbital energies and detuning. When the norm of the tunneling operator coupling these sectors is small compared with $\Delta_{\mathrm{ch}}$, their admixture probability is quadratic in this ratio to leading order; virtual energy corrections remain. The present fixed-occupation model does not determine this charge gap or the tunneling amplitudes. A quantitative implementation must evaluate them together with the geometric clearance in Eq.~\eqref{eq:clearance} and the barrier profile.

Superconducting circuit quantum electrodynamics and controlled atomic systems provide alternative settings for engineering effective quantum interactions~\cite{Blais2021,Browaeys2020}. Realizing the present model in either setting would require an explicit encoding of its orbital states, kinetic spectrum, Fourier couplings, and symmetry operations. Such an encoding would provide a route to simulating the linked-ring Hamiltonian.

\section{Conclusion}\label{sec:conclusion}
We have examined which properties of an orbital Bell pair survive coupling to higher orbitals in two Hopf-linked rings. For the ideal geometry, the Coulomb identity $V_{1,0}=-V_{0,1}$ makes the positive-parity block diagonal in the four-state Bell basis. In the full orbital Hamiltonian, the unitary symmetry $\mathcal X$ forbids mixing between opposite parities to all orders. Higher orbitals can nevertheless admix with either Bell state within its symmetry sector.

The second unitary symmetry, $\mathcal Y$, determines what this admixture means for entanglement between the rings. For every pure state in the $\mathcal Y=-1$ sector the nonzero Schmidt weights occur in equal pairs, giving $S\geq1$ bit even when all orbitals are retained. This bound also holds throughout closed evolution if the Hamiltonian preserves $\mathcal Y$ at every time. Projecting each ring coherently onto its two lowest unperturbed orbitals yields the Bell state $|\Psi_2^+\rangle$ with conditional concurrence one whenever the projection succeeds. Its success probability, $w_{\mathcal P}$, falls to about $0.57$ on the sampled branch. The angular densities characterize the spatial correlations of the global ground state; the entropy bound applies to that state when it belongs to this sector.

The robustness of these results depends on which symmetry a perturbation preserves. Equal effective-flux detunings on the two rings retain $\mathcal Y$ and its entropy bound; unequal detunings generally break that symmetry. A one-body angular potential changes the Bell-state admixture, while an elliptic deformation can preserve the Hopf link and change the Coulomb selection rule at first order. Finite ring width, tunneling and gate screening require a device-specific model. In the symmetric model, higher-orbital dressing can weaken the original two-qubit encoding without eliminating entanglement between the rings.

\appendix
\section{Spin rotation, periodicity and exchange signs}\label{app:spinsteps}
\subsection{Differential operator and spin rotation}\label{apx:spin_rotation}
We choose $-\pi/2<\gamma<\pi/2$, so $\cos\gamma>0$ and $\sqrt{1+\tan^2\gamma}=\sec\gamma$. All identities below act on differentiable spinors, with the physical spinor periodic in $\theta$. Since $p=-i\partial_\theta$,
\begin{align}
 p(\sigma_r\psi)&=-i(\partial_\theta\sigma_r)\psi+\sigma_r p\psi,\nonumber\\
 \{p,\sigma_r\}\psi&=2\sigma_rp\psi-i(\partial_\theta\sigma_r)\psi.\label{eq:equation11}
\end{align}
Using $\sigma_r^2=\mathbb I$, expansion of $[p+(\tan\gamma/2)\sigma_r]^2$ therefore retains both terms of the Hermitian Rashba operator. Write $U=AB$, with $A=e^{-i\theta\sigma_z/2}$ and $B=e^{i\gamma\sigma_y/2}$. The elementary Pauli identities give:
\begin{align}
 A^\dagger\sigma_r A&=\sigma_x,\nonumber\\
 B^\dagger\sigma_xB&=\cos\gamma\,\sigma_x-\sin\gamma\,\sigma_z,\nonumber\\
 B^\dagger\sigma_zB&=\sin\gamma\,\sigma_x+\cos\gamma\,\sigma_z.\label{eq:equation12}
\end{align}
Differentiating $A$ while $B$ is constant gives:
\begin{equation}
 -iU^\dagger\partial_\theta U
 =-\tfrac12(\sin\gamma\,\sigma_x+\cos\gamma\,\sigma_z).\label{eq:equation13}
\end{equation}
The derivative of the local spin rotation supplies the connection term in $U^\dagger pU$.

In Eq.~\eqref{eq:rashba}, the constant offset cancels the term quadratic in the Rashba coupling, leaving $4E_zp^2+[\lambda_R/(2R)]\{p,\sigma_r\}$. This retains the angular derivative of $\sigma_r$ required for Hermiticity. Since $U^\dagger pU=p-iU^\dagger\partial_\theta U$, direct differentiation gives:
\begin{align}
 U^\dagger[p+(\tan\gamma/2)\sigma_r]U
 &=p+(\tfrac12\tan\gamma\cos\gamma-\tfrac12\sin\gamma)\sigma_x\nonumber\\
 &\quad-(\tfrac12\tan\gamma\sin\gamma+\tfrac12\cos\gamma)\sigma_z \nonumber
\\
 &=p-\tfrac12\sec\gamma\,\sigma_z.\label{eq:spinsteps}
\end{align}
Because $U(\theta+2\pi)=-U(\theta)$, write the physical spinor as $U(\theta)e^{-i\theta/2}\varphi(\theta)$ with periodic $\varphi$. The phase factor shifts $p$ by $-1/2$ in the operator acting on the periodic amplitude. For the rotated spin eigenvalue $-\sigma$, the kinetic covariant momentum becomes $p-1/2+\sigma\sec\gamma/2=p-\Phi_{\mathrm{eff}}^{(j,\sigma)}/\Phi_0$, yielding Eq.~\eqref{eq:flux}.

The two minus signs acquired by $U$ and $e^{-i\theta/2}$ on a $2\pi$ circuit cancel, so $\varphi$ has integer Fourier momenta. At $\gamma=\pi/3$, the two fluxes are $3/2$ and $-1/2$, which differ by two orbital flux units. With the common Rashba energy offset omitted as in Sec.~\ref{sec:geometry}, both kinetic minima equal $E_z$. The exchange shifts $-J$ and $+J$ separate these minima by $2J$. For two electrons the common kinetic minimum is $2E_z$. Exciting one angular momentum from $1$ to $0$ or from $2$ to $3$ changes its squared covariant momentum from $1/4$ to $9/4$, at cost $8E_z$. This supplies the kinetic denominator $8E_z$ used in $\lambda$.

The aligned exchange operator satisfies $U^\dagger H_{ex}^{local}U=-J\sigma_z$. It gives $+\sigma J$ under the same branch convention. At $\gamma=\pi/3$, the desired $\sigma=-1$ orbital minimum lies at $p=1,2$ and is lower in exchange energy than the conjugate minimum. Additional SOC terms or texture misalignment that modify the projection must be included as perturbations.
Before spin projection, antiunitary invariance also depends on how the exchange texture and boundary conditions transform. Section~\ref{app:magnetic_group}, Eqs.~\eqref{eq:X}--\eqref{eq:equation9}, gives the symmetry operations and their invariance within the selected branch.

\section{Domain of the center-based multipole expansion}\label{app:multipole}
The center-based expansion in Eq.~\eqref{eq:multipole} follows from the Legendre generating function. To obtain the series, put $r=|\mathbf r_1-\mathbf r'_2|$ and $x=r/d$. Then,
\begin{equation}
 |\mathbf d+\mathbf r'_2-\mathbf r_1|^{-1}
 =d^{-1}(1-2x\cos\Theta+x^2)^{-1/2}.\label{eq:equation16}
\end{equation}
The generating function $(1-2xz+ x^2)^{-1/2}=\sum_{n\geq0}P_n(z)x^n$ for $|x|<1$ yields Eq.~\eqref{eq:multipole}. Here $z=\cos\Theta$, and $x_{\max}$ bounds the expansion ratio on the angular domain. For $x\leq x_{\max}<1$, the bound $|P_n(z)|\leq1$ for real $|z|\leq1$ majorizes the series by $\sum_nx_{\max}^n$. This justifies termwise angular integration in that domain.

For the two circles, Sec.~\ref{sec:multipole_main} gives $x_{\max}=2R/d$. The uniform bound therefore applies for $d>2R$, outside the linked domain.

\section{Explicit virtual-state correction and cancellation}\label{app:virtual}
Equations~\eqref{eq:equation10} and \eqref{eq:Qgroups} define the two nearest single-excitation groups; Eq.~\eqref{eq:VPQ} gives the coupling matrix for the upper group. The following transformation retains the full symmetry-invariant virtual space and makes the cancellation explicit.

\subsection{Generator and second-order matrix elements}\label{apx:sw_generator}
Decompose $V_{int}=V_{diag}+V_{off}$, where $V_{diag}=PV_{int}P+QV_{int}Q$ and $V_{off}=PV_{int}Q+QV_{int}P$. Choose $P\mathcal SP=Q\mathcal SQ=0$. The notation $O(V_{int}^3)$ below denotes terms cubic or higher in a formal interaction expansion; $O(\eta^n)$ elsewhere denotes the corresponding order in the perturbation amplitude. Introducing a formal order in $V_{int}$, the exponential expansion is:
\begin{align}
 e^{-\mathcal S}He^{\mathcal S}
 ={}&H_0+V_{diag}+V_{off}+[H_0,\mathcal S]\nonumber\\
 &+[V_{diag}+V_{off},\mathcal S]
 +\tfrac12[[H_0,\mathcal S],\mathcal S]+O(V_{int}^3).\label{eq:equation17}
\end{align}
The cancellation condition is:
\begin{equation}
 [H_0,\mathcal S]=-V_{off}.\label{eq:equation18}
\end{equation}
After imposing this condition, projection removes $[V_{diag},\mathcal S]$, since it is off diagonal between $P$ and $Q$. The surviving correction is $\delta H^{(2)}=\tfrac12P[V_{off},\mathcal S]P$.

Taking the generator $\mathcal S$ to be anti-Hermitian, this condition gives
$\mathcal S_{ad}=-V_{ad}/(E_a^{(0)}-E_d^{(0)})$.
Here $V_{ad}=\langle a|V_{int}|d\rangle$ and $E_a^{(0)}$ is the unperturbed energy of $|a\rangle$; $a,b$ index retained states and $d$ indexes virtual states. The second-order matrix element follows from the commutator:
\begin{align}
 \delta H_{ab}^{(2)}
 &=\tfrac12\sum_{d\in Q}(V_{ad}\mathcal S_{db}-\mathcal S_{ad}V_{db})\nonumber\\
 &=\tfrac12\sum_{d\in Q}V_{ad}V_{db}
 \left(\frac1{E_b^{(0)}-E_d^{(0)}}+
 \frac1{E_a^{(0)}-E_d^{(0)}}\right).\label{eq:SWmatrix}
\end{align}
For $a,b\in P$, $E_a^{(0)}=E_b^{(0)}=E_0$, hence $\delta H_{ab}^{(2)}=-\sum_dV_{ad}V_{db}/(E_d^{(0)}-E_0)$. The overall negative denominator sign follows from the positive excitation energies.

\subsection{Paired virtual spaces and the complete block}\label{apx:sw_pairs}
Define
\begin{equation}
 R_u=\begin{pmatrix}0&1&0&0\\1&0&0&0\\0&0&0&1\\0&0&1&0\end{pmatrix}.\label{eq:equation20}
\end{equation}
The operation $\mathcal X$ maps the ordered upper group to lower indices $(2,1,4,3)$. Invariance gives the dimensional coupling identities:
\begin{align}
 V_{\mathcal{PQ}^-}&=\Sigma_x V_{\mathcal{PQ}^+}R_u,\nonumber\\
 V_{\mathcal{PQ}^-}V_{\mathcal{PQ}^-}^\dagger
 &=\Sigma_x V_{\mathcal{PQ}^+}V_{\mathcal{PQ}^+}^\dagger\Sigma_x.
 \label{eq:virtual_pair}
\end{align}
Thus the sum of the upper and lower corrections commutes with $\Sigma_x$. Within the ordered double-excitation group, $\mathcal X$ reverses the four indices; its matrix is the same anti-diagonal identity as $\Sigma_x$. Consequently,
$\Sigma_x V_{\mathcal{PQ}^d}=V_{\mathcal{PQ}^d}\Sigma_x$,
where the left and right matrices act on the ordered $\mathcal P$ and $\mathcal Q^d$ groups, respectively. Multiplying by the adjoint shows that this group preserves parity separately. Including its doubled excitation energy, the correction from the twelve virtual states in the sixteen-state product basis is:
\begin{equation}
 \begin{aligned}
 \delta H^{(2)}={}&-\frac{V_{\mathcal{PQ}^+}V_{\mathcal{PQ}^+}^\dagger
 +V_{\mathcal{PQ}^-}V_{\mathcal{PQ}^-}^\dagger}{\Delta E_{\mathrm{kin}}}\\
 &-\frac{V_{\mathcal{PQ}^d}V_{\mathcal{PQ}^d}^\dagger}{2\Delta E_{\mathrm{kin}}}.
 \end{aligned}\label{eq:SW12}
\end{equation}
For $M_{ij}=\langle\Psi_i^+|\delta H^{(2)}|\Psi_j^-\rangle$, invariance gives $M_{ij}=\langle\Psi_i^+|\Sigma_x\delta H^{(2)}\Sigma_x|\Psi_j^-\rangle=-M_{ij}$, proving Eq.~\eqref{eq:SWzero}. In the ordered Bell basis $(\Psi_1^+,\Psi_2^+,\Psi_1^-,\Psi_2^-)$, the full correction therefore has the form
\begin{equation}
 \delta H_B^{(2)}=\begin{pmatrix}
 a&0&0&0\\0&b&0&0\\0&0&c&d\\0&0&d^*&e
 \end{pmatrix}.\label{eq:SWBell}
\end{equation}
Here $a,b,c,e$ are real matrix entries and $d$ is the off-diagonal entry in the negative block; these letters are local to this matrix and are distinct from orbital-state indices. The additional positive-block zero follows from $\mathcal Y$.

\subsection{Operator proof and range of the conclusion}\label{apx:sw_resolvent}
The four-state projector obeys $\mathcal XP=P\mathcal X$, since $p\mapsto3-p$ maps $\{1,2\}$ to itself. Hence $Q$ is invariant. Using $P\mathcal XP=\Sigma_x$, put $X_Q=Q\mathcal XQ$, $B=PV_{int}Q$ and $R_0=[Q(H_0-E_0)Q]^{-1}$. The symmetry identities are
$\Sigma_xB=BX_Q, [X_Q,R_0]=0, X_Q^2=Q$.
Therefore,
\begin{align}
 \Sigma_x\delta H^{(2)}\Sigma_x
 &=-\Sigma_xBR_0B^\dagger \Sigma_x\nonumber\\
 &=-BX_QR_0X_QB^\dagger=-BR_0B^\dagger=\delta H^{(2)}.\label{eq:equation21}
\end{align}
Equivalently, for $A_{id}=\langle\Psi_i^+|V_{int}|d\rangle$ and $B_{dj}=\langle d|V_{int}|\Psi_j^-\rangle$, a partner $\bar d=\mathcal Xd$ obeys
$A_{i\bar d}=A_{id}, B_{\bar d j}=-B_{dj}, E_{\bar d}^{(0)}=E_d^{(0)}$.
Its contribution cancels that of $d$ in every opposite-parity entry. If a virtual eigenvector itself has definite parity, one of the two amplitudes vanishes instead. Retaining an upper group without its lower partner violates the hypothesis and can create a spurious matrix element.

The same structural conclusion can be checked without a series expansion. Write an exact eigenvector as $\psi_P+\psi_Q$, where $\psi_P=P\psi$, $\psi_Q=Q\psi$, and $E$ is its eigenenergy. Whenever $E-QHQ$ is invertible on $Q$, its $Q$ equation gives $\psi_Q=(E-QHQ)^{-1}QHP\psi_P$. Substitution into the $P$ equation yields:
\begin{equation}
 H_{eff}(E)=PHP+PHQ(E-QHQ)^{-1}QHP.\label{eq:exact_resolvent}
\end{equation}
Since $[P,\mathcal X]=[Q,\mathcal X]=0$ and $[H,\mathcal X]=0$, the restricted operator $QHQ$ and its resolvent commute with the restriction of $\mathcal X$ to $Q$. The two coupling operators in Eq.~\eqref{eq:exact_resolvent} intertwine its restrictions to $P$ and $Q$. Consequently $[H_{eff}(E),\Sigma_x]=0$ on $\mathcal P$, with $\Sigma_x=\mathcal X|_{\mathcal P}$. The same argument applies to $\mathcal Y$. This is an exact statement on the resolvent domain; each coefficient of a symmetry-preserving perturbative expansion is likewise block diagonal. Convergence of that expansion is a separate requirement. At a pole the elimination formula is unavailable, while the full Hamiltonian still commutes with $\mathcal X$. Orbital dressing and energy shifts within a sector are thus compatible with the absence of opposite-parity mixing to all orders.

\begin{table}[t]\caption{Signed contributions to $\langle\Psi_1^+|\delta H^{(2)}|\Psi_1^-\rangle$ in units $E_c^2/(8E_z)$ for $\alpha=0.3$, $\beta=1.1$. Adaptive quadrature uses requested absolute and relative tolerances $10^{-8}$. The residuals indicate numerical cancellation.}\label{tab:SW}
\begin{ruledtabular}\begin{tabular}{ccc}
Virtual group&$\rho_{\mathrm{soft}}=0$&$\rho_{\mathrm{soft}}=0.05$\\
Upper&$+0.011583$&$+0.011539$\\
Lower&$-0.011583$&$-0.011539$\\
All 12, magnitude&$2.4\times10^{-18}$&$1.4\times10^{-18}$
\end{tabular}\end{ruledtabular}\end{table}
Table~\ref{tab:SW} reports the signed virtual-group contributions. Real and imaginary Fourier components are integrated independently, without imposing the analytical symmetry relations. All Bell-matrix entries are evaluated at five quadrature tolerances $10^{-4},\ldots,10^{-8}$ and compared with a calculation at tolerance $10^{-10}$. Independent periodic quadrature reproduces this twelve-virtual-state correction to $1.9\times10^{-15}$. The cancellation residual is already at floating-point precision and need not decrease monotonically with quadrature tolerance. The finite virtual-state calculation is consistent with the operator identity, which applies to the full symmetry-invariant virtual space.

\section{Numerical methods and reproducibility}\label{app:numerics}
All orbital calculations use the unexpanded kernel of Sec.~\ref{sec:coulomb} and real Fourier coefficients for the stated ideal geometries. Periodic quadrature samples equally spaced angles with the endpoint omitted; Fourier-transform normalization is $1/N^2$. The distance is evaluated as a sum of squares:
\begin{align}
 \mathcal D^2={}&(\cos\theta_1-\beta-\cos\theta_2)^2\nonumber\\
 &+(\sin\theta_1+\sin\alpha\sin\theta_2)^2
 +\cos^2\alpha\sin^2\theta_2,\label{eq:Dstable}
\end{align}
which avoids subtractive cancellation near close approaches. A constant $\rho_{\mathrm{soft}}^2$ is added only where specified. Angular refinement and orbital refinement are separate checks. Successive-grid differences provide empirical convergence estimates. These estimates quantify stability under grid refinement rather than absolute integration errors.

We write $V_{m,n}^{(N)}$ for the finite-grid approximation to $V_{m,n}$. Here $N$ is the number of angular points on each ring, and $K$ is the orbital-window cutoff defined below. Applying periodic quadrature to Eq.~\eqref{eq:fourier} on $\theta_{j,k}=2\pi k/N$ gives
\begin{equation}
 \frac{V_{m,n}^{(N)}}{E_c}=\frac1{N^2}\sum_{k,l=0}^{N-1}
 \frac{e^{2\pi i(mk+nl)/N}}{\mathcal D(2\pi k/N,2\pi l/N)}.\label{eq:equation22}
\end{equation}
This is the positive-phase inverse discrete Fourier transform. Negative transfer indices use their residues modulo $N$ for access to the quadrature array; the orbital Hamiltonian itself has no such wraparound. For a fixed $K$, its dimensionless entries are given below, with $\delta_{pp'}$ denoting the Kronecker delta (one for equal indices and zero otherwise). In the matrix subscripts, primes distinguish column-state indices from row-state indices. The entries are:
\begin{align}
 (H/(4E_z))_{pq,p'q'}={}&[(p-3/2)^2+(q-3/2)^2]\nonumber\\
 &\times\delta_{pp'}\delta_{qq'}+2\lambda V_{p'-p,q'-q}/E_c.\label{eq:equation23}
\end{align}
The maximum needed transfer is $2K+1$. Angular refinement compares all used transfers; orbital refinement then tests the observables obtained from the corresponding Hamiltonian. These two checks address different approximations.

The orbital window $p=1-K,\ldots,2+K$ has $(2K+2)^2$ product states and is closed under $\mathcal X$ and $\mathcal Y$. Every momentum-transfer entry of the interaction is retained. Sector diagonalization uses orthonormal projectors and is checked against full-matrix calculations at representative points. Pure-state entropy is obtained from squared singular values of $C$, independently checked against eigenvalues of $CC^\dagger$. Individual eigenstates are identified when their separation from neighboring eigenvalues exceeds the numerical resolution. The angular grids, orbital cutoffs and observable-specific convergence criteria are given below.

For Figs.~\ref{fig:symmetry} and \ref{fig:energy_analysis}, coefficients are refined on the stated $N$ sequence until each dimensionless coefficient changes by at most $10^{-9}+10^{-7}|V_{m,n}|/E_c$. Figure~\ref{fig:concurrence} also requires concurrence changes below $10^{-8}+10^{-7}|\mathcal C_-|$, with exact zero-tilt odd integrals treated by their analytical symmetry. Figure~\ref{fig:mixing} uses a softened kernel, coefficient tolerance $10^{-9}+10^{-7}|V_{m,n}|/E_c$, angular grids $N=256,512,1024,2048,4096$, separately evaluated tilt slices and adaptive-quadrature checks of selected points. For the softening-dependent maximum in Fig.~\ref{fig:mixing}, nested $31\times31$ and $61\times61$ grids sample $0\leq\alpha\leq1$ and $0.01\leq\beta\leq1.99$ at each of the 20 softening values. Refining the geometry grid changes the maximum by at most $3.12\times10^{-7}$. The refined maxima decrease from $0.14838$ at $\rho_{\mathrm{soft}}=0.001$ to $0.096048$ at $\rho_{\mathrm{soft}}=0.2$. Independent adaptive checks at the maximizing sampled geometries for three softening values agree with the nine corresponding coefficients within $5.7\times10^{-14}$. The resulting maximum refers to the sampled geometries; values between grid points are not constrained by these comparisons.

\begin{table}[tbp]
\caption{Orbital-cutoff convergence at $\alpha=0.3$ rad, $\beta=1.1$, $\lambda=0.03$, and $\rho_{\mathrm{soft}}=0$. The orbital window $p=1-K,\ldots,2+K$ contains $(2K+2)^2$ product states. The upper block gives the internal splitting of the two positive-parity Bell-connected branches and its absolute change from the preceding listed cutoff, in units of $E_c$. The lower block gives the four-state projection deficit $1-w_{\mathcal P}$, entropy above the one-bit lower bound, $S-1$ (with $S$ in bits), and conditional concurrence $\mathcal C_{\mathcal P}$ of the $\Psi_2^+$ branch. Column headings specify scale factors. Successive changes are evaluated before rounding; the last two are given as upper limits on the observed changes.}
\label{tab:cutoff}
\centering\small
\setlength{\tabcolsep}{3pt}
\begin{tabular}{rrrr}
\hline\hline
$K$ & States & $10^2\Delta E_+/E_c$ & Successive change \\
\hline
0 & 4 & 4.3242 & -- \\
1 & 16 & 4.3126 & $1.17\times10^{-4}$ \\
2 & 36 & 4.3123 & $2.82\times10^{-6}$ \\
3 & 64 & 4.3123 & $1.99\times10^{-7}$ \\
4 & 100 & 4.3123 & $1.86\times10^{-8}$ \\
6 & 196 & 4.3123 & $2.25\times10^{-9}$ \\
8 & 324 & 4.3123 & $<10^{-10}$ \\
10 & 484 & 4.3123 & $<10^{-11}$ \\
\hline\hline
\end{tabular}
\par\smallskip
\begin{tabular}{rrrr}
\hline\hline
$K$ & $10^5(1-w_{\mathcal P})$ & $10^7(S-1)$ & $\mathcal C_{\mathcal P}$ \\
\hline
0 & 0 & 0 & 1 \\
1 & 3.0438 & 5.3814 & 1 \\
2 & 3.0774 & 5.7861 & 1 \\
3 & 3.0785 & 5.9136 & 1 \\
4 & 3.0786 & 5.9352 & 1 \\
6 & 3.0786 & 5.9387 & 1 \\
8 & 3.0786 & 5.9388 & 1 \\
10 & 3.0786 & 5.9388 & 1 \\
\hline\hline
\end{tabular}
\end{table}
Table~\ref{tab:cutoff} shows convergence of the Bell-connected splitting to approximately $0.043123E_c$. The change from $K=8$ to $K=10$ is below $10^{-11}E_c$. The global lowest excitation gap, a distinct quantity, approaches $0.23064E_c$. For the $\Psi_2^+$ branch, the corresponding changes in projection weight and full entropy are below $10^{-13}$ and $10^{-12}$ bits, respectively. The conditional concurrence remains unity within numerical precision because the normalized four-state projection retains its Bell-state form in this symmetry sector. The projection weight and entropy separately quantify orbital dressing and its cutoff dependence. These comparisons establish empirical convergence at the stated parameter point; the parameter-dependent tests for the remaining figures are specified below. Fourier coefficients use $N=512$ after comparison with $N=256$; independent adaptive-quadrature checks use absolute tolerance $10^{-10}$ and relative tolerance $10^{-9}$.

For Fig.~\ref{fig:fullentropy}, the absolute cutoff criteria are $10^{-5}$ bits in entropy, $10^{-6}$ in weight, $10^{-6}\Delta E_{\mathrm{kin}}$ in energy and $10^{-5}\Delta E_{\mathrm{kin}}$ in the within-sector gap. For all 400 parameter points, the changes from $K=8$ to $K=10$ satisfy these tolerances. Overlaps between adjacent interaction points additionally check branch continuity. Fourier coefficients satisfy an absolute-plus-relative comparison of $10^{-9}+10^{-7}|V_{m,n}|/E_c$; accepted angular grids are $N=400$ or $3200$.

For Fig.~\ref{fig:localization}, Fourier coefficients are refined on $N=150,300,600,1200$ until the largest transfer-coefficient change is below $10^{-11}$. Spread and energy changes must be below $10^{-6}$ and $(10^{-7})(4E_z)$. All 237 points on the three spread curves meet these criteria at $K=10$ and $N=300$. The two density maps additionally require an integrated absolute density change below $10^{-5}$, maximum pointwise change below $10^{-5}$ and gap change below $(10^{-6})(4E_z)$. Both density maps satisfy these criteria upon increasing the cutoff from $K=8$ to $K=10$. The coordinate-space circular moments also agree with the coefficient-space result to numerical precision. Cell-centered plotting aligns the angular samples with image boundaries.

For Fig.~\ref{fig:robustness}, observable changes in conditional concurrence, fidelity, entropy and weight must be below $10^{-5}$, with energy and within-sector gap changes below $(10^{-6})(4E_z)$ and $(10^{-5})(4E_z)$, respectively. The accepted cutoffs range from $K=10$ to $20$, with angular grids $N=400$ or $800$. These tolerances are satisfied at all 62500 parameter points. The material/entropy map in Fig.~\ref{fig:phasediagram} uses coefficient tolerances $10^{-9}+10^{-7}|V_{m,n}|/E_c$ and absolute cutoff tolerances $10^{-5}$ bits in entropy, $10^{-6}$ in weight and $(10^{-6})(4E_z)$ in both ground-state energy and gap. For all 2500 parameter points, the changes from $K=8$ to $K=10$ satisfy the stated tolerances.

\section{Shape asymmetry, angular potentials and flux detuning}\label{app:imperfections}
A topology-preserving deformation can change the Fourier identity because that identity requires a particular geometric bijection. To see the first-order dependence, perturb the dimensionless relative displacement $\mathbf D$ by $\eta\,\delta\mathbf D$. At fixed unperturbed angular measure and basis,
\begin{equation}
 \delta V_{m,n}=-\frac{E_c}{(2\pi)^2}\int\!\!\int
 e^{i(m\theta_1+n\theta_2)}
 \frac{\mathbf D\cdot\delta\mathbf D}{\mathcal D^3}\,
 d\theta_1d\theta_2.\label{eq:shapevariation}
\end{equation}
For the softened kernel the denominator becomes $(\mathcal D^2+\rho_{\mathrm{soft}}^2)^{3/2}$. An actual shape model also changes the line element, kinetic operator, and possibly the local SOC texture; those contributions must accompany Eq.~\eqref{eq:shapevariation}. If the resulting perturbation retains $\mathcal X$ but breaks $\mathcal Y$, the sum $\delta V_{1,0}+\delta V_{0,1}$ enters the positive-block off-diagonal element. Perturbations with a parity-odd component additionally couple the positive and negative sectors through Eq.~\eqref{eq:paritypert}.

The coefficient variation follows by differentiating the kernel itself:
\begin{align}
 \left.\frac{d}{d\eta}|\mathbf D+\eta\delta\mathbf D|^{-1}\right|_{\eta=0}
 &=-\frac12(\mathbf D^2)^{-3/2}\,2\mathbf D\cdot\delta\mathbf D\nonumber\\
 &=-\frac{\mathbf D\cdot\delta\mathbf D}{\mathcal D^3}.\label{eq:equation24}
\end{align}
At positive minimum separation this differentiation can be passed through the compact angular integral for bounded smooth deformations sufficiently small to retain separation. Equation~\eqref{eq:shapevariation} denotes $\partial_\eta V_{m,n}|_0$, so the actual first-order change is $\eta\,\delta V_{m,n}$.

As a concrete Coulomb-kernel example, stretch only ring 1 along $x$:
$\mathbf r_1^{(\eta)}/R=((1+\eta)\cos\theta_1,\sin\theta_1,0)$,
with ring 2 unchanged. For sufficiently small $|\eta|$ the centerlines remain disjoint and linked. In Eq.~\eqref{eq:shapevariation}, $\delta\mathbf D=(\cos\theta_1,0,0)$, giving the first-order change in the identity as
\begin{equation}
\begin{aligned}
 \Gamma(\eta)&=\frac{V_{1,0}(\eta)+V_{0,1}(\eta)}{E_c},\\
 \Gamma'(0)&=-\left\langle
 \frac{D_x\cos\theta_1(\cos\theta_1+\cos\theta_2)}{\mathcal D^3}
 \right\rangle,\\
 D_x&=\cos\theta_1-\beta-\cos\theta_2.
\end{aligned}
\label{eq:elliptic_derivative}
\end{equation}
In Eq.~\eqref{eq:elliptic_derivative}, $\langle\cdot\rangle$ denotes the uniform average over both angles. The imaginary parts vanish under simultaneous angular inversion. At $(\alpha,\beta)=(0.3,1.1)$ and $\rho_{\mathrm{soft}}=0$, direct periodic quadrature gives $0.20408$ for this derivative in the fixed-angle Fourier convention. A centered finite difference at $\eta=\pm10^{-3}$ differs from the unrounded derivative by $3.45\times10^{-8}$. Periodic quadrature on $N=128,256,512,768$ angular points per ring gives the same derivative to the displayed precision. The nonzero derivative demonstrates that this smooth, topology-preserving deformation changes the Fourier identity at first order. The result isolates the Coulomb-kernel contribution in the fixed-angle representation. Predicting entanglement in an elliptic-ring device additionally requires arc-length normalization, the deformed kinetic and spin--orbit operators, and the corresponding orbitals.

The potential $V_{dis}=\epsilon E_c\cos\theta_1$ of Sec.~\ref{sec:robustness} has projected off-diagonal amplitude $\epsilon E_c/2$ because $\langle p|\cos\theta|p\pm1\rangle=1/2$. After subtracting the common diagonal term, the positive Bell block is
\begin{equation}
 \begin{pmatrix}W&\epsilon E_c/2\\\epsilon E_c/2&-W\end{pmatrix},\label{eq:equation25}
\end{equation}
with eigenvalues $\pm\sqrt{W^2+(\epsilon E_c/2)^2}$. Within this four-state model, the two eigenvalue surfaces form a conical intersection in the $(W,\epsilon E_c/2)$ plane. The direction of approach to the origin determines the limiting eigenvectors. For its normalized real eigenvectors, the coefficient difference $|c_1^2-c_2^2|$ gives Eq.~\eqref{eq:Cpert}. At $W=0$ and $\epsilon\ne0$ the eigenvectors are $(\Psi_1^+\pm\Psi_2^+)/\sqrt2$, which factor into equal-sign or opposite-sign local superpositions. If $W=\epsilon=0$, the entire block is degenerate and the eigenvector requires a separate selection prescription. In this four-state model the fidelity to the corresponding unperturbed Bell eigenvector is:
\begin{equation}
 F_+^{(4)}=\tfrac12\left[1+\frac{|W|}{\sqrt{W^2+(\epsilon E_c/2)^2}}\right],\qquad W\ne0.\label{eq:Fpert}
\end{equation}
The enlarged-orbital computation in Fig.~\ref{fig:robustness} supplies the corresponding dressed-state quantities.

For the general perturbation in Eq.~\eqref{eq:paritypert}, write $|\psi\rangle=|a,+\rangle+\eta|\psi^{(1)}\rangle+O(\eta^2)$ and $E=E_a+\eta E_a^{(1)}+O(\eta^2)$. The first-order eigenvalue equation is:
\begin{equation}
 (H-E_a)|\psi^{(1)}\rangle=-(A-E_a^{(1)})|a,+\rangle.\label{eq:equation26}
\end{equation}
Taking its overlap with $\langle b,-|$ gives the energy denominator in Eq.~\eqref{eq:paritypert}. Since the parity-even part of $A$ has zero matrix element between opposite sectors, only $A_o$ contributes. For an isolated eigenvalue, the leading opposite-parity probability is:
\begin{equation}
 \|P_-\psi\|^2=\eta^2\sum_{b,-}
 \frac{|\langle b,-|A_o|a,+\rangle|^2}{(E_{a,+}-E_{b,-})^2}
 +O(\eta^3),\label{eq:equation27}
\end{equation}
where $P_-=(1-\mathcal X)/2$. A lower bound $g_{opp}>0$ on all coupled opposite-parity energy differences bounds the leading coefficient by $\|P_-A_o|a,+\rangle\|^2/g_{opp}^2$. The perturbative regime requires the leading probability in Eq.~\eqref{eq:equation27} to be much smaller than unity. If one opposite-parity level dominates and diagonal perturbative shifts are negligible, substantial two-level mixing occurs when $2|\eta\langle b,-|A_o|a,+\rangle|$ becomes comparable to $|E_{a,+}-E_{b,-}|$. Diagonal shifts must be included in the level separation when significant. This criterion is specific to the coupled states; the internal positive-parity splitting is a different quantity.

A local effective-flux error $\Phi_{\mathrm{eff}}^{(j)}/\Phi_0=3/2+\delta f_j$ produces:
\begin{equation}
 \delta H_{flux}=-8E_z\sum_j\delta f_j(p_j-3/2)+4E_z\sum_j\delta f_j^2.\label{eq:fluxerror}
\end{equation}
Its linear term is odd under $\mathcal X$. For equal detunings, $\delta f_1=\delta f_2=\delta f$, this term is $-8E_z\delta f(p_1+p_2-3)$ and is invariant under $\mathcal Y$, which exchanges $p_1$ and $p_2$. The quadratic term shifts all orbital energies by the same constant. Since the interaction is also $\mathcal Y$-invariant, the detuned orbital Hamiltonian commutes with $\mathcal Y$ for any common detuning. Thus mixing of opposite $\mathcal X$ sectors can coexist with the one-bit entropy bound for pure $\mathcal Y=-1$ states proved in Appendix~\ref{apx:entropy_bound}. Within $\mathcal P$, the $\mathcal Y=-1$ eigenspace is spanned solely by $\Psi_2^+$, so its nonzero projection remains a Bell state even without a definite $\mathcal X$ label. The encoding weight remains state dependent, and a common detuning generally lifts the original kinetic degeneracy.

For unequal detunings, the linear term additionally contains $-4E_z(\delta f_1-\delta f_2)(p_1-p_2)$, which is odd under $\mathcal Y$. The sector argument then gives no general entanglement bound. These conclusions assume that flux errors enter through Eq.~\eqref{eq:fluxerror}; additional changes of the spin texture or scalar potential require their own symmetry analysis. Static shape disorder and time-dependent noise must likewise be specified through their spatial form, strength and, for noise, temporal statistics before an error rate or lifetime can be inferred.

\section{Full-space unitary and antiunitary operations}\label{app:operatorsteps}
\subsection{Hilbert space, domains and coordinate actions}\label{apx:domains}
The symbol $\mathbb T^2$ denotes the two-dimensional torus of periodic angles, $\mathbb Z$ the integers, and $L^2$ the square-integrable wavefunctions. The orbital Hilbert space for the periodic amplitudes of Appendix~\ref{apx:spin_rotation} is $L^2(\mathbb T^2,d\theta_1d\theta_2)$ with orthonormal functions $\langle\theta_1,\theta_2|p,q\rangle=e^{i(p\theta_1+q\theta_2)}/(2\pi)$, $p,q\in\mathbb Z$. The kinetic operator has the periodic second-derivative domain. At fixed disjoint geometry the Coulomb multiplication operator is bounded, since $\min\mathcal D>0$, so the full Hamiltonian has the same domain. The transformations below preserve periodicity and this domain. They can first be checked on finite Fourier sums and then extended to the operator domain.

For $f$ in this space, define
\begin{align}
 (If)(\theta_1,\theta_2)&=f(-\theta_1,-\theta_2),\nonumber\\
 (\Pi_{12}f)(\theta_1,\theta_2)&=f(\pi-\theta_2,\pi-\theta_1),\nonumber\\
 (Mf)(\theta_1,\theta_2)&=e^{3i(\theta_1+\theta_2)}f(\theta_1,\theta_2).\label{eq:equation28}
\end{align}
All angles are understood modulo $2\pi$. The first two maps preserve angular measure; $M$ multiplies by a unit-modulus periodic factor. Thus they are unitary. Substitution gives
$I^2=\Pi_{12}^2=1, [I,\Pi_{12}]=0, IMI=M^{-1}$.
It follows that $\mathcal X=MI$ and $\mathcal Y=I\Pi_{12}$ are involutions. In coordinates,
\begin{align}
 (\mathcal Xf)(\theta_1,\theta_2)&=e^{3i(\theta_1+\theta_2)}f(-\theta_1,-\theta_2),\nonumber\\
 (\mathcal Yf)(\theta_1,\theta_2)&=f(\theta_2+\pi,\theta_1+\pi).\label{eq:equation29}
\end{align}
The phase $M$ is unchanged by $\mathcal Y$, because its extra phase is $e^{6\pi i}=1$; hence $[\mathcal X,\mathcal Y]=0$.

\subsection{Conjugation of momentum and interaction}\label{apx:conjugation}
Applying the chain rule to the coordinate actions in Appendix~\ref{apx:domains} gives:
\begin{align}
 Ip_jI&=-p_j,&Mp_jM^{-1}&=p_j-3,\nonumber\\
 \Pi_{12}p_1\Pi_{12}^{-1}&=-p_2,&
 \Pi_{12}p_2\Pi_{12}^{-1}&=-p_1.\label{eq:equation30}
\end{align}
Therefore $\mathcal Xp_j\mathcal X^{-1}=M(-p_j)M^{-1}=3-p_j$, while $\mathcal Y$ exchanges $p_1$ and $p_2$. Squaring the shifted momenta proves:
\begin{align}
 \mathcal X(p_j-3/2)^2\mathcal X^{-1}&=(p_j-3/2)^2,\nonumber\\
 \mathcal YH_0\mathcal Y^{-1}&=H_0.\label{eq:equation31}
\end{align}
The latter equality uses equal kinetic coefficients and equal effective fluxes on the rings. Since $M$ commutes with every multiplicative potential, $\mathcal XV_{int}\mathcal X^{-1}=IV_{int}I=V_{int}$. Invariance under $I$ and $\Pi_{12}$ gives the corresponding identity for $\mathcal Y$. Thus both the kinetic and interaction terms are invariant under these operations.

\subsection{Antiunitarity and the commutator}\label{apx:antiunitarity}
Let $(K_cf)(\theta_1,\theta_2)=f(\theta_1,\theta_2)^*$. Then $K_c(af+bg)=a^*K_cf+b^*K_cg$ and $\langle K_cf|K_cg\rangle=\langle g|f\rangle$. The antiunitary operator $\mathcal K_{\mathrm{global}}^{\mathrm{orb}}=K_c\Pi_{12}$ consequently acts as:
\begin{equation}
 (\mathcal K_{\mathrm{global}}^{\mathrm{orb}}f)(\theta_1,\theta_2)=f(\pi-\theta_2,\pi-\theta_1)^*.\label{eq:equation32}
\end{equation}
The real coordinate substitution commutes with complex conjugation, so $(\mathcal K_{\mathrm{global}}^{\mathrm{orb}})^2=K_c^2\Pi_{12}^2=1$. Direct application to a basis wave gives:
\begin{align}
 \mathcal K_{\mathrm{global}}^{\mathrm{orb}}\frac{e^{i(p\theta_1+q\theta_2)}}{2\pi}
 &=\frac{e^{-i[p(\pi-\theta_2)+q(\pi-\theta_1)]}}{2\pi}\nonumber\\
 &=(-1)^{p+q}\frac{e^{i(q\theta_1+p\theta_2)}}{2\pi}.\label{eq:equation33}
\end{align}
Together with antilinearity, this proves Eq.~\eqref{eq:antiunitary}. Momentum conjugation gives $\mathcal K_{\mathrm{global}}^{\mathrm{orb}}p_1(\mathcal K_{\mathrm{global}}^{\mathrm{orb}})^{-1}=K_c(-p_2)K_c^{-1}=p_2$, and similarly for $p_2$. Thus $\mathcal K_{\mathrm{global}}^{\mathrm{orb}}H_0(\mathcal K_{\mathrm{global}}^{\mathrm{orb}})^{-1}=H_0$ at equal flux and equal kinetic coefficient. Reality and parity-exchange invariance of $V_{int}$ give $\mathcal K_{\mathrm{global}}^{\mathrm{orb}}V_{int}(\mathcal K_{\mathrm{global}}^{\mathrm{orb}})^{-1}=V_{int}$.

In coefficient space, denote componentwise conjugation by $K_b$; it differs from coordinate-space $K_c$, which also reverses the Fourier momenta. The antiunitary representation is $\mathcal K_{\mathrm{global}}^{\mathrm{orb}}=Y K_b$, with $Y$ the real matrix of $\mathcal Y$. The matrix condition is $YH^*Y^{-1}=H$. For transfers $m=p'-p$, $n=q'-q$, writing $V$ for the matrix of $V_{int}$, its interaction entry is:
\begin{align}
 (YV^*Y^{-1})_{pq,p'q'}
 &=(-1)^{p+q+p'+q'}V_{q'-q,p'-p}^*\nonumber\\
 &=(-1)^{m+n}V_{n,m}=V_{m,n}.\label{eq:equation34}
\end{align}
Here reality and the Fourier identity are used at the last step. This supplies an independent coefficient-space version of the coordinate calculation. Because this antiunitary operation squares to the identity, a nondegenerate eigenstate can be chosen invariant under $\mathcal K_{\mathrm{global}}^{\mathrm{orb}}$. The parity quantum number used here belongs to the linear unitary $\mathcal X$.

\subsection{Restriction to the encoding and joint sectors}\label{apx:joint_sectors}
In the ordered product basis of $\mathcal P$, the restriction of $\mathcal X$ is $\Sigma_x$ from Eq.~\eqref{eq:sigmax_main}. The matrix of $P\mathcal YP$ is
\begin{equation}
 Y_P=\begin{pmatrix}1&0&0&0\\0&0&-1&0\\0&-1&0&0\\0&0&0&1\end{pmatrix}.\label{eq:equation35}
\end{equation}
On $(\Psi_1^+,\Psi_2^+,\Psi_1^-,\Psi_2^-)$ their diagonal entries are respectively $(+,+,-,-)$ and $(+,-,+,+)$. Any Hermitian operator commuting with both has the Bell-block form in Eq.~\eqref{eq:SWBell}. In a symmetry-closed enlarged window, $x,y\in\{+1,-1\}$ denote the eigenvalues of $\mathcal X,\mathcal Y$; the joint-sector projector is:
\begin{equation}
 P_{xy}=\tfrac14(1+x\mathcal X)(1+y\mathcal Y),\qquad x,y=\pm1.\label{eq:equation36}
\end{equation}
Commuting involutions give $P_{xy}^2=P_{xy}$, orthogonality for different labels and $\sum_{x,y}P_{xy}=1$. These identities justify the sector diagonalizations and their state labels.

\section{Bell-block spectra and entanglement observables}\label{app:observable_steps}
\subsection{Explicit basis transformation and sector spectra}\label{apx:bell_spectra}
The columns of the orthogonal change-of-basis matrix
\begin{equation}
 B=\frac1{\sqrt2}\begin{pmatrix}
 1&0&1&0\\0&1&0&1\\0&1&0&-1\\1&0&-1&0
 \end{pmatrix}\label{eq:equation37}
\end{equation}
are the four ordered Bell states. Subtracting the common energy $E_0$ gives $B^\dagger(H_{\mathcal P}-E_0\mathbb I)B=H^+\oplus H^-$, with the blocks in Eq.~\eqref{eq:blocks}. For example,
\begin{align}
 \langle\Psi_1^+|V_{int}|\Psi_2^+\rangle
 &=V_{0,1}+V_{1,0}=0,\nonumber\\
 \langle\Psi_1^-|V_{int}|\Psi_2^-\rangle
 &=\tfrac12(V_{0,1}-V_{1,0}-V_{-1,0}+V_{0,-1})\nonumber\\
 &=V_{0,1}-V_{1,0}=-2V_{1,0}.\label{eq:equation38}
\end{align}
The positive-block diagonal entries are $V_{0,0}+V_{1,1}$ and $V_{0,0}+V_{1,-1}$. Opposite-parity entries vanish from $[H_{\mathcal P},\Sigma_x]=0$. Subtracting the mean of the negative block leaves:
\begin{equation}
 \begin{aligned}
 h_-&=\begin{pmatrix}-W&-2V_{1,0}\\-2V_{1,0}&W\end{pmatrix},\\
 \det(h_- -z\mathbb I)&=z^2-W^2-4V_{1,0}^2.
 \end{aligned}\label{eq:equation39}
\end{equation}
Here $h_-$ is the negative block after subtraction of its mean, and $z$ is the spectral variable in its characteristic polynomial. The two roots are $\pm\sqrt{W^2+4V_{1,0}^2}$, giving the sector splittings stated in Sec.~\ref{app:bell}. The eigenvalues of the positive-parity block relative to its mean are $\pm W$; taking their separation gives $2|W|$ independently of level ordering.

\subsection{Concurrence and exceptional cases}\label{apx:bell_concurrence}
A parity state has coefficient matrix:
\begin{equation}
 C_+=\frac1{\sqrt2}\begin{pmatrix}c_1&c_2\\c_2&c_1\end{pmatrix},\qquad
 C_-=\frac1{\sqrt2}\begin{pmatrix}c_1&c_2\\-c_2&-c_1\end{pmatrix}.\label{eq:equation40}
\end{equation}
Normalization is $|c_1|^2+|c_2|^2=1$. The pure-state formula $2|\det C_\pm|$, where $\det$ denotes the determinant, yields $|c_1^2-c_2^2|$, including complex coefficients. For the real negative-block eigenvectors, the spectral projector of $h_-=\mathbf h\cdot\bm\tau$, $\mathbf h=(-2V_{1,0},0,-W)$, with $\bm\tau=(\tau_x,\tau_y,\tau_z)$ the Pauli matrices in this two-dimensional block, is $(1\pm h_-/|\mathbf h|)/2$. Hence $|c_1^2-c_2^2|=|W|/\sqrt{W^2+4V_{1,0}^2}$. This formula requires $W^2+4V_{1,0}^2>0$. If both vanish, the negative block is degenerate and concurrence depends on the selected vector. Similarly, the positive-block result one refers to its separate Bell eigenvectors when $W\ne0$; at $W=0$ the selection must be specified.

\subsection{Partial trace, projection weight and fidelity}\label{apx:partial_trace}
For the normalized state in Sec.~\ref{sec:fullentropy}, $\sum_{pq}|c_{pq}|^2=1$. Taking the trace over ring 2 explicitly gives:
\begin{align}
 \varrho_1&=\sum_q {}_2\langle q|\psi\rangle\langle\psi|q\rangle_2,\nonumber\\
 (\varrho_1)_{pp'}&=\sum_qc_{pq}c_{p'q}^*=(CC^\dagger)_{pp'}.\label{eq:equation41}
\end{align}
If $C=U_C\operatorname{diag}(s_k)V_C^\dagger$ is a singular-value decomposition, with unitary matrices $U_C,V_C$, nonnegative singular values $s_k$ and $\operatorname{diag}$ denoting a diagonal matrix, the reduced-state eigenvalues are $s_k^2$, with $\sum_ks_k^2=1$. Thus $S=-\sum_ks_k^2\log_2s_k^2$, with $0\log_20=0$. A rank-one product state has $S=0$; a Bell state has two equal eigenvalues $1/2$ and $S=1$ bit.

Projection retains $c_{11},c_{12},c_{21},c_{22}$, whose squared moduli sum to $w_{\mathcal P}$. Dividing the retained coefficient matrix by $\sqrt{w_{\mathcal P}}$ divides its determinant by $w_{\mathcal P}$, proving Eq.~\eqref{eq:conditional}. At $w_{\mathcal P}=0$ conditional concurrence is undefined. In the $(\mathcal X,\mathcal Y)=(+1,-1)$ sector, the projection has the form $P\psi=a\Psi_2^+$ because that joint sector is one dimensional within $P$. Consequently $w_{\mathcal P}=|a|^2=|\langle\Psi_2^+|\psi\rangle|^2$ and the conditional concurrence is one when $a\ne0$. This identity concerns the projected encoding, while the singular values of $C$ determine the entanglement between the rings.

\subsection{Entropy bound in the \texorpdfstring{$\mathcal Y=-1$}{Y=-1} sector}\label{apx:entropy_bound}
For a state with $\mathcal Y|\psi\rangle=-|\psi\rangle$, Eq.~\eqref{eq:Y} gives
$c_{pq}=-(-1)^{p+q}c_{qp}$. Let $D$ be the diagonal unitary matrix $D_{pp}=(-1)^p$, and set $C_{\mathrm a}=DC$. Then $(C_{\mathrm a})_{pq}=-(C_{\mathrm a})_{qp}$, so $C_{\mathrm a}^T=-C_{\mathrm a}$, while $C_{\mathrm a}C_{\mathrm a}^\dagger=D\varrho_1D^\dagger$ has the same spectrum as $\varrho_1$. For a finite complex antisymmetric matrix, the positive singular values have even multiplicity (Corollary~2 of Ref.~\cite{Youla1961}). One can see this directly: if $C_{\mathrm a}C_{\mathrm a}^\dagger u=s^2u$ with $s>0$, then $v=C_{\mathrm a}\bar u/s$ is an orthogonal eigenvector with the same eigenvalue. Indeed, $u^\dagger C_{\mathrm a}\bar u=0$ by antisymmetry, $C_{\mathrm a}^\dagger C_{\mathrm a}=(C_{\mathrm a}C_{\mathrm a}^\dagger)^*$, and the antilinear map $u\mapsto C_{\mathrm a}\bar u/s$ squares to $-1$ in this eigenspace. Hence every nonzero eigenvalue of $\varrho_1$ appears an even number of times. Because the eigenvalues sum to one, none exceeds $1/2$, and
\begin{equation}
 S=-\sum_k s_k^2\log_2 s_k^2\geq1\ \mathrm{bit}.
\label{eq:entropy_lower_bound}
\end{equation}
The argument holds in every symmetry-closed finite orbital window. For a normalizable full-space state, $C_{\mathrm a}$ is a Hilbert--Schmidt operator, meaning that the sum of the squared magnitudes of its matrix entries is finite. It is therefore compact, and each positive eigenvalue of $C_{\mathrm a}C_{\mathrm a}^\dagger$ has a finite-dimensional eigenspace to which the same pairing argument applies. Equation~\eqref{eq:entropy_lower_bound} thus extends to the full ring-mode space, allowing infinite entropy. Equality is obtained for one nonzero equal-weight pair. The bound applies to pure states in the $\mathcal Y=-1$ sector. Their overlap with $\Psi_2^+$ is determined separately by the projection weight; states outside this sector require their own entanglement analysis.

This bound also constrains closed-system dynamics. Let $U_{\mathrm{orb}}(t,0)$ denote the unitary propagator generated by $H(t)$ on the orbital Hilbert space, assuming the evolution is well defined. If $[H(t),\mathcal Y]=0$ at every time, then $\mathcal Y U_{\mathrm{orb}}(t,0)\mathcal Y$ satisfies the same Schr\"odinger equation and initial condition as $U_{\mathrm{orb}}(t,0)$. Uniqueness therefore gives $[U_{\mathrm{orb}}(t,0),\mathcal Y]=0$. An initial pure state in the $\mathcal Y=-1$ sector remains in that sector and satisfies $S(t)\geq1$ bit at every time. The argument includes all orbitals and requires no adiabatic or weak-coupling approximation. It bounds the entropy without fixing its value or the projection weight. Every term introduced by a physical driving protocol must preserve the same symmetry for this conclusion to apply; general dissipative evolution is outside this pure-state statement.

\subsection{Angular normalization and relative circular moment}\label{apx:angle_moment}
For the angular density of Sec.~\ref{sec:localization}, Fourier orthogonality gives:
\begin{equation}
 \int_{-\pi}^{\pi}\!\int_{-\pi}^{\pi}|\psi|^2\,d\theta_1d\theta_2
 =\sum_{p,q}|c_{pq}|^2=1.\label{eq:equation42}
\end{equation}
Multiplication of a Fourier basis function by $e^{i(\theta_1-\theta_2)}$ raises $p$ by one and lowers $q$ by one. Inserting the expansion on both sides of the expectation value therefore leaves precisely $\sum_{pq}c_{p+1,q-1}^*c_{pq}$. Coefficients outside the retained window are zero in this finite-state evaluation; no wraparound is imposed on the orbital indices. Since multiplication by this phase is unitary, $|Z_{\mathrm{rel}}|\leq1$ and $-\ln|Z_{\mathrm{rel}}|^2\geq0$ when finite. Distinct angular distributions can share the same first harmonic; the joint density maps show spatial structure that this moment alone cannot distinguish.

\bibliographystyle{apsrev4-2}
\bibliography{references}
\end{document}